\documentclass[12pt,a4paper]{article} 

\usepackage{amsmath,amssymb,cite,euscript}
\usepackage{epsfig,psfrag, subfigure}
\usepackage{color}
\usepackage{bm}
\begin{document} 

\begin{titlepage}

\begin{flushright}
MZ-TH/09-32\\ [0.2cm]
September 9, 2009 \\  
\end{flushright}

\vspace*{5mm}
\begin{center}
    {\baselineskip 25pt
    \Large\bf

\boldmath{NNLO corrections to $\bar B \to X_u \ell \bar \nu_\ell$ and the
    determination of $|V_{ub}|$} 
    }

\vspace{0.8cm}
\begin{center}
{\sc C.~Greub$^{(a)}$, M.~Neubert$^{(b)}$, and 
B.~D.~Pecjak$^{(b)}$}\\
\vspace{0.7cm}
$^{(a)}${\sl  Albert Einstein Center for Fundamental Physics,
    Institute for Theoretical Physics, Univ. of Bern, CH-3012 Bern,
Switzerland
 } \\
\vspace{0.3cm}
$^{(b)}${\sl Johannes Gutenberg-Universit\"at, D-55099 Mainz, Germany} \\
\vspace{0.3cm}
\end{center}

\end{center}

\bigskip

\centerline{\bf Abstract}

\medskip 

\noindent 
We study the impact of next-to-next-to-leading order  (NNLO) QCD corrections
on partial decay rates in $\bar B \to X_u \ell \bar \nu_\ell$ decays,
at leading-order in the $1/m_b$ expansion for shape-function
kinematics. These corrections are implemented within a modified form
of the BLNP framework, which allows for arbitrary variations of 
the jet scale $\mu_i\sim 1.5$~GeV.  Our analysis includes a 
detailed comparison between resummed and  fixed-order perturbation 
theory, and between the complete NNLO results and
those obtained in the large-$\beta_0$ approximation.
For the default choice $\mu_i=1.5$~GeV used in current extractions
of $|V_{ub}|$ within the BLNP framework, the NNLO 
corrections induce significant downward shifts in the 
central values of partial decay rates with cuts on the hadronic 
variable $P_+$, the hadronic invariant mass, and the lepton energy.     
At the same time, perturbative uncertainties are reduced, especially
those at the jet scale, which are the dominant ones at next-to-leading
order (NLO).  For higher values of $\mu_i$ and in 
fixed-order perturbation theory, the shifts between NLO and NNLO are 
more moderate. We combine our  new results with known 
power-suppressed terms in order to illustrate the implications of 
our analysis on the determination of $|V_{ub}|$ from inclusive decays.

\end{titlepage}

\newpage

\section{Introduction} 
\label{sec:Introduction}
The CKM element $|V_{ub}|$  is a fundamental parameter of 
flavor physics.  It measures the strength of $b\to u$ quark 
transitions and determines the length of the side of the unitarity
triangle opposite to the angle $\beta$.  The combined work
of many theorists and experimentalists has allowed to determine this 
parameter from data on both inclusive and exclusive 
semi-leptonic $b\to u$ decays;  a summary of current results 
can be found in, for instance, \cite{Barberio:2008fa, Antonelli:2009ws,
Neubert:2008cp}.  
A characteristic feature of these analyses is that the value of 
$|V_{ub}|$ deduced from inclusive 
decays is consistently higher than that from exclusive decays.
The need to resolve this discrepancy motivates systematic
improvements on all fronts.

The goal of this paper is to improve the theory predictions 
for the inclusive decays  $\bar B \to X_u \ell \bar \nu_\ell$
by including recently calculated next-to-next-to-leading order
(NNLO) perturbative corrections to the partial decay rates
used in the extraction of $|V_{ub}|$.  In general, these partial
rates are available only in the portion of phase space
referred to as the shape-function region, where the hadronic final
state is a jet carrying an energy of order $m_b$ and an invariant
mass squared of order $m_b\Lambda_{\rm QCD}$. A local 
operator product expansion is not valid in this region, and results 
are obtained through a factorization formalism, as a  
convolution of perturbative hard-scattering kernels with 
non-perturbative shape functions 
\cite{Neubert:1993ch, Neubert:1993um, Bigi:1993ex}.   
Different approaches to this 
factorization have been put forth in the literature, going under 
the names of  BLNP \cite{Bosch:2004th,Lange:2005yw},  
GGOU \cite{Gambino:2007rp}, and the dressed-gluon 
exponentiation \cite{Andersen:2005mj}.  

In this paper we implement the NNLO perturbative corrections within the 
BLNP framework. In this approach,  techniques from soft-collinear
effective theory (SCET) \cite{Bauer:2000yr,Bauer:2001yt, Beneke:2002ph} 
are used to obtain an arbitrary partial decay 
rate as an expansion in $1/m_b$, according to 
\begin{equation}
\label{eq:mbexp}
\Gamma_u=\Gamma_u^{(0)}+[\Gamma_u^{(1)}+\Gamma_u^{(2)}+\dots ].
\end{equation}
The first term is of leading order in the heavy-quark 
expansion, whereas the terms in the square brackets account for 
power-suppressed effects and are estimated up to order $1/m_b^2$. 
The NNLO corrections 
studied in the current paper affect only the leading-order 
term $\Gamma_u^{(0)}$.  For kinematic cuts limited to the 
shape-function region, this term obeys a factorization formula
of the schematic form 
\cite{Korchemsky:1994jb, Akhoury:1995fp,Bauer:2003pi, Bosch:2004th}  
\begin{equation}
\label{eq:VagueFact}
\Gamma_u^{(0)} \sim [H\cdot J](\mu_f)\otimes S(\mu_f) ,
\end{equation}
where the symbol $\otimes$ denotes a convolution.  The factorization
formula contains a hard function $H$, related to physics 
at the hard scale $\mu_h\sim m_b$, a jet function $J$,  related 
to physics at the intermediate scale 
$\mu_i \sim (m_b\Lambda_{\rm QCD})^{1/2}\sim 1.5$~GeV, 
and a non-perturbative shape function $S$, describing 
the internal soft dynamics of the $B$ meson.     
A thorough analysis of this leading-power term at next-to-leading
order (NLO) in renormalization-group (RG) improved 
perturbation theory  was performed in 
\cite{Lange:2005yw}.  We can extend this analysis to NNLO by
putting together a number of pieces, which we describe in  
Section~\ref{sec:PartialRates}. In addition to calculating
the higher-order perturbative corrections, we also 
modify the BLNP framework to allow for variations of 
the arbitrary matching scale $\mu_i$ at which the jet function is 
calculated. This allows us to study the perturbative uncertainties
associated with this scale, and makes for a straightforward matching 
with fixed-order perturbation theory, where logarithms between 
the hard and intermediate scales are not resummed.
In the numerical analysis in
Section \ref{sec:Numerics} we apply our new results to 
partial decay rates with cuts on the hadronic variable $P_+$, the 
hadronic invariant mass, and the lepton energy.  
Our main findings are that the dependence on the scale $\mu_i$ 
is sizeable at NLO and still significant even at NNLO, and that for 
the default choice $\mu_i=1.5$~GeV typically used in the BNLP 
framework the NNLO corrections  tend to shift the partial 
decay rates downward by a significant amount.
We also study some qualitative aspects of the perturbative series 
by comparing results obtained in resummed and fixed-order 
perturbation theory, and with those obtained in 
the large-$\beta_0$ approximation. To illustrate the significance
of our results for the extraction of $|V_{ub}|$, in Section \ref{sec:Vub}
we combine the NNLO corrections to the leading-order term
with known power corrections and experimental data in order to
extract sample values of $|V_{ub}|$ for several partial rates.  
For the particular choice of intermediate scale 
$\mu_i=2.0$~GeV,  the effect of the NNLO corrections is to 
raise the central value of $|V_{ub}|$ by slightly less than 
10\% compared to the results at NLO; for higher choices of the 
intermediate scale and in fixed-order perturbation theory, the corrections
are more moderate.
We summarize our findings in Section \ref{sec:Conclusions}.

\section{Partial decay rates in SCET} 
\label{sec:PartialRates}
In this section we briefly review the BLNP formalism as applied 
to inclusive semi-leptonic $b\to u$ decays.  
We begin by recalling the master formula for an arbitrary partial
decay rate derived in \cite{Lange:2005yw}, expressed in terms of
the hadronic variables
\begin{equation}
 P_- = E_X + |\vec P_X| \,, \qquad 
 P_+ = E_X - |\vec P_X| \,.
\end{equation}
It is given by
\begin{equation}
\label{eq:maspar}
   \frac{d\Gamma_u(y_{\rm max},y_0)}{dP_+}
   = \left\{ \begin{array}{lll}
   \Gamma_u^A(y_{\rm max}) &;& \quad y_{\rm max} \le y_0 \,, \\
   \Gamma_u^A(y_0) + \Gamma_u^B &;& \quad y_{\rm max} > y_0 \,,
   \end{array} \right.
\end{equation}
where 
\begin{eqnarray}
   \Gamma_u^A(y_i)
   &=& \frac{G_F^2|V_{ub}|^2}{96\pi^3}\,(M_B-P_+)^5\,
    \int_0^{y_i}\!dy\,y^{2} \left[ (3-2y)\,f_1
    + 6(1-y)\,f_2 + y\,f_3 \right] , \nonumber \\
   \Gamma_u^B
   &=& \frac{G_F^2|V_{ub}|^2}{96\pi^3}\,(M_B-P_+)^5\,
    \int_{y_0}^{y_{\rm max}}\!dy\, y_0\\
   &&\times \left[ \left( 6y(1+y_0) - 6y^2
    - y_0(3+2y_0) \right) f_1
    + 6y(1-y)\,f_2
    + y_0(3y-2y_0)\,f_3 \right]  \nonumber.
\end{eqnarray}
The dimensionless variables $y$, $y_0$, and $y_{\rm max}$ are defined as
\begin{equation}
 y = \frac{P_--P_+}{M_B-P_+} \,, \qquad  y_{\rm max} = 
\frac{P_-^{\rm max}-P_+}{M_B-P_+} \,, \qquad
   y_0 = \frac{P_l^{\rm max}-P_+}{M_B-P_+} = 1 - \frac{2 E_0}{M_B-P_+} \,,
\end{equation}
and we have introduced the quantity $E_0$, which is the minimum allowed
lepton energy.  The values of $y_{\rm max}$ and the integration range for 
$P_+$ depend on the specifics of the partial decay rate under consideration.
For cuts $P_+<\Delta_P$ and on the lepton energy, we have $y_{\rm max}=1$ and 
$0<P_+ < {\rm min}(\Delta_P,M_B-2E_0)$.   For a cut on the
hadronic invariant mass, $M_X<M_0$,  we have
\begin{equation}
y_{\rm max}=\frac{{\rm min}(M_B,M_0^2/P_+)-P_+}{M_B-P_+}
\end{equation}
and $0<P_+ < M_0$.

The scalar functions $f_i$ are obtained as an expansion in 
$1/m_b$.  The leading-order term has the form (\ref{eq:VagueFact}) and 
reads
\begin{equation}\label{eq:BuLP}
   f_i^{\rm (0)}(P_+,y)
   = H_{ui}(y,m_b,\mu_f) \int_0^{P_+}\!d\hat\omega\,y m_b\,
    J(y m_b(P_+-\hat\omega),\mu_f)\,\hat S(\hat\omega,\mu_f) \,.
\end{equation}
As written, (\ref{eq:BuLP}) achieves a factorization of 
perturbative and non-perturbative physics.  The hard 
functions $H_{ui}$ contain physics at the scale $m_b$,
the jet function $J$ contains physics at the jet scale
$(m_b\Lambda_{\rm QCD})^{1/2}$, and the shape-function $S$ is 
a non-perturbative matrix element in heavy-quark effective
theory (HQET) \cite{Neubert:1993ch, Bigi:1993ex}.  In SCET, one 
usually assumes the parametric limit where the hard scale
is much larger than the jet scale, in which case 
any choice of $\mu_f$ leads to large perturbative logarithms in 
either $H$ or $J$.
To solve this problem one first calculates these functions at 
matching scales where they contain no 
large logarithms, and then evolves them to a common scale 
$\mu_f$ using the renormalization group (RG).
Also, within the effective-theory framework, it is natural to
extract the non-perturbative shape function at a low
scale $\mu_0$, and then  evolve it to the scale $\mu_f$.  
The RG evolution can be achieved using results from 
\cite{Bosch:2004th, Neubert:2004dd} 
for the hard and shape functions, and \cite{Becher:2006nr} 
for the jet function.  For convenience, we list the solutions here:
\begin{eqnarray}
\label{eq:resummedHJS}
H_{ui}(y,m_b, \mu_f) &=&
y^{-2a_\Gamma(\mu_h,\mu_f)} {\rm exp}\left[2 S(\mu_h,\mu_f)-2 a_\Gamma(\mu_h,\mu_f)\ln\frac{m_b}{\mu_h}-2a_{\gamma^\prime}(\mu_h,\mu_f)\right] \nonumber \\
&& \hspace{2cm} \times H_{ui}(y,m_b,\mu_h)\,,  \\ && \nonumber \\
 J(p^2,\mu_f) &=& \exp\left[ - 4S(\mu_i,\mu_f) + 2 a_{\gamma^J}(\mu_i,\mu_f) \right]
   \widetilde j(\partial_{\eta_J},\mu_i) \left[ \frac{1}{p^2}
   \left( \frac{p^2}{\mu_i^2} \right)^{\eta_J} \right]_{\!*}\,
   \frac{e^{-\gamma_E\eta_J}}{\Gamma(\eta_J)}, \nonumber \\ &&  \nonumber \\
\hat S(\hat \omega,\mu_f)&=& {\rm exp}
\left[2 S(\mu_0,\mu_f)+2 a_{\gamma^{'}-\gamma^{J}}(\mu_0,\mu_f)\right]\,
\frac{e^{-\gamma_E \eta_S}}{\Gamma(\eta_S)}\int_0^{\hat \omega}
d\hat \omega' \frac{\hat S(\hat \omega',\mu_0)}
{\mu_0^{\eta_S}(\hat \omega -\hat \omega')^{1-\eta_S}}. \nonumber
\end{eqnarray}
The RG exponents read \cite{Bosch:2004th}
\begin{equation}
\label{eq:RGexps}
   S(\nu,\mu)
   = -\int\limits_{\alpha_s(\nu)}^{\alpha_s(\mu)}\!\!d\alpha\,
    \frac{\Gamma_{\rm cusp}(\alpha)}{\beta(\alpha)} 
    \int\limits_{\alpha_s(\nu)}^{\alpha}\frac{d\alpha'}{\beta(\alpha')}
    \,, \qquad
   a_\Gamma(\nu,\mu)
   = -\int\limits_{\alpha_s(\nu)}^{\alpha_s(\mu)}\!\!d\alpha\,
    \frac{\Gamma_{\rm cusp}(\alpha)}{\beta(\alpha)} \,,
\end{equation}
and similarly for $a_{\gamma'}$ ($a_{\gamma^J}$), but with 
$\Gamma_{\rm cusp}$ replaced by the anomalous dimensions of 
the hard (jet) function.  We have defined  $\eta_J = 2 a_\Gamma(\mu_i,\mu_f)$, 
 $\eta_S=2a_\Gamma(\mu_f,\mu_0)$, and
$\widetilde j$ is the Laplace transform of $J$.  Finally, 
the star distribution is defined as 
\begin{equation}\label{star}
   \int_0^{Q^2}\!dp^2\,\left[ \frac{1}{p^2}
   \left( \frac{p^2}{\mu^2} \right)^\eta \right]_{\!*}\,f(p^2)
   = \int_0^{Q^2}\!dp^2\,\frac{f(p^2)-f(0)}{p^2}
    \left( \frac{p^2}{\mu^2} \right)^\eta
    + \frac{f(0)}{\eta} \left( \frac{Q^2}{\mu^2} \right)^\eta .
\end{equation}
Inserting the above results into the
factorization formula (\ref{eq:BuLP}), and using the general
relations \cite{Becher:2006mr}
\begin{eqnarray}
   a_\Gamma(\mu_1,\mu_2) + a_\Gamma(\mu_2,\mu_3)
   &=& a_\Gamma(\mu_1,\mu_3) \,, \nonumber\\
   S(\mu_1,\mu_2) + S(\mu_2,\mu_3)
   &=& S(\mu_1,\mu_3) + \ln\frac{\mu_1}{\mu_2}\,a_\Gamma(\mu_2,\mu_3) \,,
\end{eqnarray}
one  is left with
\begin{eqnarray}\label{eq:Resummedf}
&& f_i^{(0)}(P_+,y)= {\rm exp}\bigg[2 S(\mu_h,\mu_i)-2S(\mu_i,\mu_0)
+2 a_{\gamma^J}(\mu_i,\mu_0)-2a_{\gamma'}(\mu_h,\mu_0) 
\nonumber \\
&& 
\hspace{3cm}- 2a_{\Gamma}(\mu_h,\mu_i)\ln\frac{m_b}{\mu_h}\bigg] 
\times H_{ui}(y,m_b,\mu_h)\,y^{-2a_\Gamma(\mu_h,\mu_i)}  \\
&&\hspace{3cm}\times
 \widetilde j\left(\ln\frac{m_by}{\mu_i}+\partial_\eta,\mu_i\right)\frac{e^{-\gamma_E\eta}}{\Gamma(\eta)}
\int_0^{P_+}d\hat\omega\left[\frac{1}{P_+-\hat\omega}
\left(\frac{P_+-\hat\omega}{\mu_i}\right)^\eta\right]_* \hat S(\hat
\omega,\mu_0)\, \nonumber ,
\end{eqnarray}
where now $\eta=2 a_\Gamma(\mu_i,\mu_0)$.  Note that the dependence
on the arbitrary factorization scale $\mu_f$ has disappeared.

Equations (\ref{eq:maspar}) and (\ref{eq:Resummedf}) are the master 
formulas for calculating the leading-power contribution to 
a given partial decay rate in RG-improved 
perturbation theory.  The results are formally independent of the 
matching scales $\mu_h$ and $\mu_i$, but a residual dependence 
remains when truncating the perturbative expansion at a given 
order.  Moreover, by setting $\mu_h=\mu_i=\mu$,
one recovers fixed-order perturbation theory directly from the 
resummed results. The product of
matching functions $(H\cdot J)(\mu)$ is combined into a single coefficient
function $C(\mu)$, and the RG exponents serve to evolve
the shape-function from the scale $\mu_0$ to $\mu$.   We 
shall discuss and compare 
the perturbative uncertainties in both resummed and fixed-order 
perturbation theory at LO, NLO, and NNLO in the following section.
The scale $\mu_0$ plays a special role in the 
analysis.  We shall choose this as the scale at which we 
model the non-perturbative shape function; 
more details will be given below.
  
We can evaluate the master formula 
(\ref{eq:Resummedf}) at NNLO in RG-improved perturbation theory by 
gathering together a number of results available in the literature.    
By NNLO, we mean the approximation which captures all of the 
order $\alpha_s^2$ terms in both the matching functions and 
the RG exponents. For the matching functions 
$H_{ui}$ and $\widetilde j$ this counting is unambiguous: 
they are both needed at two loops.  The $H_{ui}$ to 
this order can be derived from the calculations in 
\cite{Bonciani:2008wf, Asatrian:2008uk, Beneke:2008ei, Bell:2008ws},
and $\widetilde j$ was calculated to NNLO in 
\cite{Becher:2006qw}; for the convenience of the reader, 
we list the results in the Appendix.
As for the RG exponents, 
to define their NNLO expansion requires
assumptions about the  matching scales $\mu_h,\mu_i$, and 
$\mu_0$. By default, we shall assume the hierarchy 
$\mu_h \gg \mu_i\gg \mu_0$.  In that case, to account for all 
of the $\alpha_s^2$ pieces in the RG exponents requires
$\Gamma_{\rm cusp}$ to four loops and the anomalous dimensions
$a_{\gamma^J}$ and $a_{\gamma'}$ to three loops.  
However, both $\gamma '$ and  $\Gamma_{\rm  cusp}$ 
are known to one loop lower, which adds a small uncertainty to 
the analysis. For the missing pieces, we shall use the [1,1] Pad\'e
approximation described in the Appendix. In practice, we shall always
use $\mu_0=1.5$~GeV in next section, so that $\mu_i\sim \mu_0$.  In
that case the general expression (\ref{eq:Resummedf}) resums some higher-order 
logarithms which are not large.

To evaluate the partial decay rates requires a model for the 
shape function $\hat S(\hat \omega,\mu_0)$.  Experimental 
information on the shape function is provided by data on the 
photon energy spectrum in $\bar B\to X_s \gamma$ decays, and can 
be used to guide the functional form of the model.
In addition, model-independent constraints are provided by the 
fact that moments of the shape function, defined as 
\cite{Bosch:2004th, Lange:2005yw}
\begin{equation}
\label{eq:Smom}
M_N(\hat \omega_0,\mu)=\int_0^{\hat \omega_0}d\hat \omega \, \hat \omega^N
\hat S(\hat \omega,\mu),
\end{equation}
can be calculated in a local heavy-quark expansion, as long 
as $\hat \omega_0\gg \Lambda_{\rm QCD}$ \cite{Bosch:2004th, Neubert:2004sp}.
In the renormalization scheme referred to as the shape-function
scheme \cite{Bosch:2004th, Neubert:2004sp}, the shape-function moments are 
used to define the heavy-quark parameters order-by-order 
in perturbation theory.  Limiting ourselves to the first two moments,
which is sufficient to order $1/m_b^2$ in the heavy-quark expansion, one has 
\begin{eqnarray}\label{MomentRelations}
   \frac{M_1(\mu_f+\bar\Lambda(\mu_f,\mu),\mu)}
        {M_0(\mu_f+\bar\Lambda(\mu_f,\mu),\mu)}
   &=& \bar\Lambda(\mu_f,\mu) \,, \nonumber\\
   \frac{M_2(\mu_f+\bar\Lambda(\mu_f,\mu),\mu)}
        {M_0(\mu_f+\bar\Lambda(\mu_f,\mu),\mu)}
   &=& \frac{\mu_\pi^2(\mu_f,\mu)}{3} + \bar\Lambda^2(\mu_f,\mu) \,,
\end{eqnarray}
where $\bar \Lambda = M_B - m_b$ and $\mu_\pi^2$ is related to the 
kinetic-energy parameter in HQET.  Requiring that a given shape-function
model correctly reproduces the moment relations (\ref{MomentRelations})
puts constraints on its parameters.  These constraints depend 
on the order in perturbation theory at which the moments are 
evaluated.  In the numerical analysis, we shall always use the 
constraints obtained from the two-loop moments.  The moments to 
this order can be calculated as explained in \cite{Neubert:2004sp},
using the two-loop expression for the quantity $s(L,\mu)$ obtained 
in  \cite{Becher:2005pd}.  In addition to the perturbative 
expressions, one needs numerical values for $\bar \Lambda(\mu_f,\mu)$
and $\mu_\pi^2(\mu_f,\mu)$.  These  
can be determined from global fits for the HQET parameters
in other renormalization schemes 
using perturbative conversion relations.  For the choice of 
scales $\mu_f=\mu=\mu_*$, the connection
with the pole scheme reads \cite{Neubert:2004sp}  
\begin{eqnarray}\label{simplebeauty}
   m_b^{\rm pole} 
   &=& m_b(\mu_*,\mu_*) + \mu_*\,\frac{4\alpha_s(\mu_*)}{3\pi}
    \left[ 1 + \frac{\alpha_s(\mu_*)}{\pi}\,
    \left( \frac{271}{36} + \frac{7\pi^2}{36} - \frac{17}{12}\,\zeta_3
     - \frac{47}{54}\,n_f \right) \right] \nonumber\\
   &&\mbox{}+ \frac{\mu_\pi^2(\mu_*,\mu_*)}{\mu_*}
    \left( \frac{\alpha_s(\mu_*)}{\pi} \right)^2
    \left( -\frac{13}{81} + \frac{17}{27}\,\zeta_3 
    + \frac{10}{243}\,n_f \right) ,\nonumber\\
   - \lambda_1^{\rm pole}
   &=& \mu_\pi^2(\mu_*,\mu_*) \left[ 1 - \frac{2\alpha_s(\mu_*)}{3\pi}
    + \left( \frac{\alpha_s(\mu_*)}{\pi} \right)^2 
    \left( \frac{19}{18} - \frac{7\pi^2}{54} - \frac{17}{6}\,\zeta_3
     - \frac{n_f}{9} \right) \right] \nonumber\\
   &&\mbox{}+ \mu_*^2 \left( \frac{\alpha_s(\mu_*)}{\pi} \right)^2
    \left( - \frac{20}{9} + \frac{17}{6}\,\zeta_3 
    + \frac{14}{27}\,n_f \right) .
\end{eqnarray}
In the rest of the paper, we will use  the notation
$m_b(\mu_*,\mu_*)\equiv m_b^*$ and  $\mu_\pi^2(\mu_*,\mu_*)\equiv \mu_\pi^{*2}$ 
for the HQET parameters in the shape-function scheme, evaluated
at the scale $\mu_*=1.5$~GeV.  Recent HFAG numbers for 
these parameters are  $m_b^*=\left(4.707^{+0.059}_{-0.053}\right)$~GeV and 
$\mu_{\pi}^{*2}=\left(0.216^{+0.054}_{-0.076}\right)$~GeV$^2$  
\cite{Barberio:2008fa}. 
They are determined from information on $\bar B \to X_c l \bar \nu_l$
moments alone, under the assumption that moments from 
$B\to X_s \gamma$ decays should not be used in the global fits,
on grounds that the measurements are typically made at values of the 
photon energy where shape-function effects are expected to be 
non-negligible \cite{Neubert:2008cp}.  To obtain numerical values
for the non-diagonal parameters  $m_b(\mu_f,\mu)$ and 
$\mu_\pi^{2}(\mu_f,\mu)$, one uses the generalization of 
(\ref{simplebeauty}) given in  \cite{Neubert:2004sp}, along with 
the fact that the pole scheme parameters are scale independent.  

\begin{figure}[t]
\begin{center}
\begin{tabular}{lr}
\psfrag{x}[]{ $ \hat \omega$ (GeV)}
\psfrag{left}[]{\raisebox{0.75cm}{
 ${\scriptsize \hat S(\hat \omega,\mu_0)}$}}
\psfrag{prelim}[]{${\scriptsize^{\mu_\pi^{*2}=0.2~{\rm GeV}^2 ,\, \, \, \mu_0=1.5~{\rm GeV}}}$ }
\psfrag{lo}[]{}
\psfrag{nlo}[]{}
\psfrag{nnlo}[]{}
\includegraphics[width=0.40\textwidth]{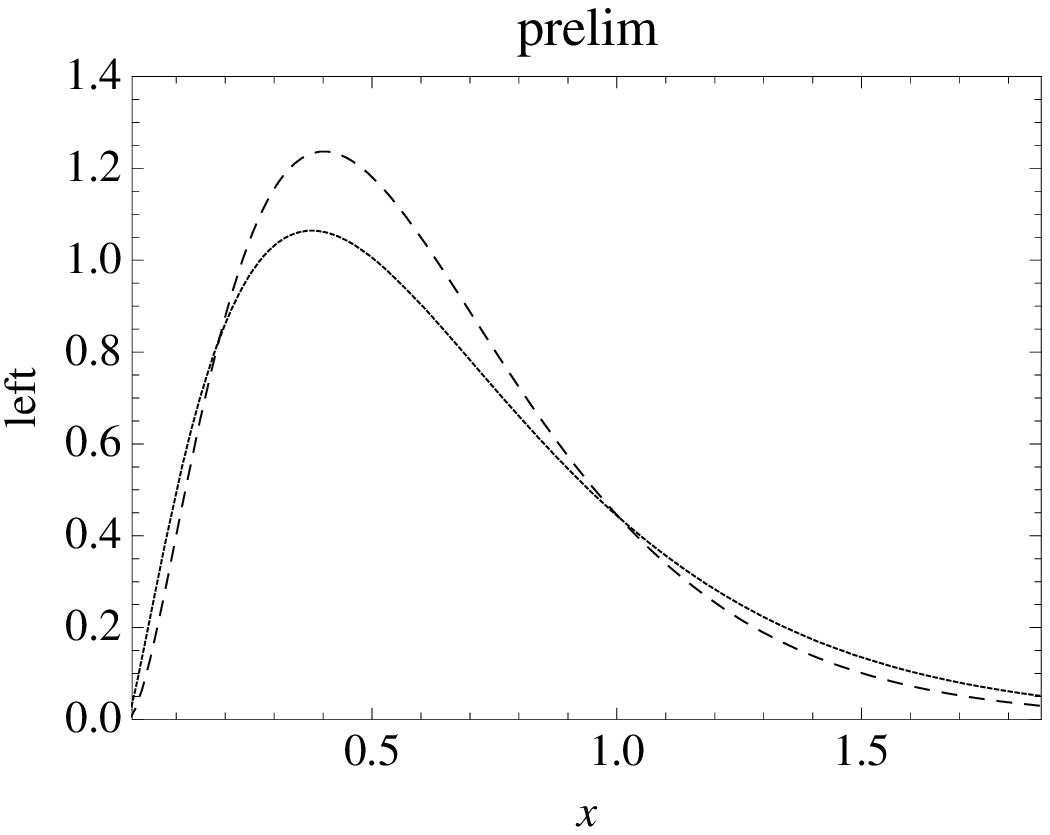}  & \hspace{1.2cm}
\psfrag{x}[]{$\hat \omega$ (GeV)}
\psfrag{left}[]{\raisebox{0.75cm}{
 ${\scriptsize \hat S(\hat \omega,\mu_0)}$}}
\psfrag{prelim}[]{${\scriptsize^{ \mu_\pi^{*2}=0.2~{\rm GeV}^2,\, \, \, \mu_0=1.5~{\rm GeV}}}$}
\psfrag{lo}[]{}
\psfrag{nlo}[]{}
\psfrag{nnlo}[]{}
\includegraphics[width=0.40\textwidth]{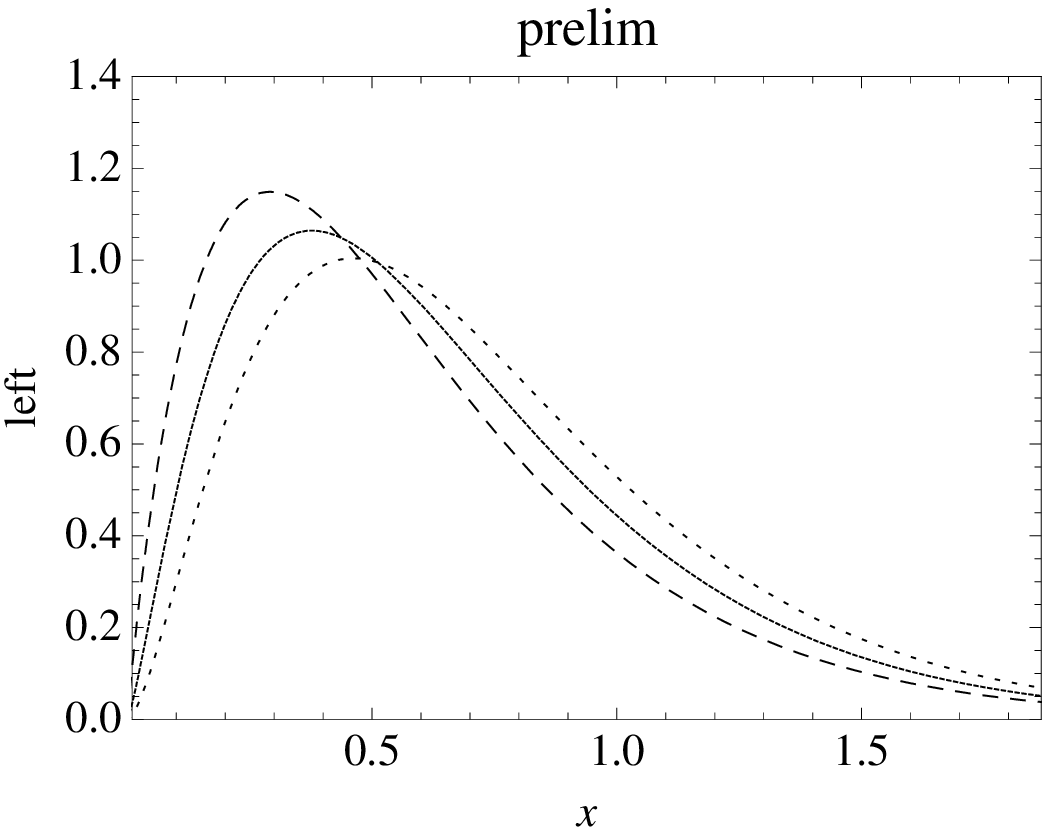}
\end{tabular}
\end{center}
\vspace{-0.5cm}
\caption{\label{fig:SFs}  
Left: model of the shape-function with the moment relations evaluated at 
one-loop order (dashed) and two-loop order (solid), 
for  $m_b^*=4.71$~GeV.  Right:
model of the shape-function with the moment relations evaluated at 
two loops, for $m_b^*=4.66$~GeV (dotted),  $m_b^*=4.71$~GeV (solid),
and  $m_b^*=4.77$~GeV (dashed).  }
\end{figure}

In the analysis that follows, we model the shape function as 
\begin{equation}
\label{eq:Smod}
\hat S(\hat \omega,\mu_0)={\cal N}(b,\Lambda)\,\hat \omega^{b-1} 
\, {\rm exp}\left(-\frac{b\hat \omega}{\Lambda}\right)  .
\end{equation}
The normalization factor ${\cal N}$, as well as the model parameters
$b$ and $\Lambda$, are tuned to satisfy the two-loop moment
constraints described above.  To get a feeling for how this 
compares with the model generated from the one-loop moments
and used in  \cite{Lange:2005yw}, we show in the left-hand 
plot of Figure~\ref{fig:SFs} the shape-function  model (\ref{eq:Smod}) 
tuned to reproduce  the moment relations for $m_b^*=4.71$~GeV 
and $\mu_{\pi}^{*2}=0.2$~GeV$^2$ at one- and two-loop order.   
In both cases, we use the four-loop running coupling 
with $\alpha_s(M_z)=0.1176$, and match onto the 
four-flavor theory at 4.25 GeV. The results also depend on 
the UV cutoff of the moment 
integration range, which we choose as $\hat \omega_0=M_B-2E_0$,
with $E_0=1.8$~GeV.  We furthermore  take $\mu_0=1.5$ GeV.  
The figure shows that the shape-function model generated using the two-loop
moment relations has a smaller average value for $\hat \omega < 1.0$~GeV 
compared to that generated with the one-loop moments. 
Since to leading order in $\alpha_s$ a given 
partial rate is directly proportional to the integral of 
the shape function over a window specified by the cut, the model with 
the two-loop constraints tends to yield lower values for partial decay rates 
with kinematics restricted to the shape-function region.  
In the right-hand plot of Figure~\ref{fig:SFs} we show the shape-function model 
generated using the NNLO moment relations for three different 
values of $m_b^*$, which cover the range of the HFAG values quoted
above.  The models with higher $m_b^*$ have a noticeably larger average 
value for $\hat \omega \lesssim 0.5$~GeV, but the difference 
starts to become smaller for values of $\hat \omega$ higher than this.  
This implies that raising
$m_b^*$  tends to raise partial decay rates in the shape-function region,
and that this effect is largest 
for the most restrictive cuts. We shall see in the next section 
that the changes in the shape-function model for different values of 
$m_b^*$ constitute the largest  parametric uncertainty in the $|V_{ub}|$ 
analysis. 

Since the emphasis of this paper is the study of the 
perturbative series for a given shape function model, in the numerical
studies we limit ourselves to the shape-function model 
(\ref{eq:Smod}), with the moment constraints implemented as explained
above.  Recently, a different procedure for building shape-function 
models which satisfy the moment constraints was proposed in 
\cite{Ligeti:2008ac}. It would be interesting to see to what extent 
the numerical results change when using such models at common values of 
the HQET parameters and the scale $\mu_0$, but we do not explore
this issue in the current work.

\section{Numerical results for leading-power partial rates}
\label{sec:Numerics}

\begin{figure}[t]
\begin{center}
\begin{tabular}{lr}
\psfrag{x}[]{$\mu_h$ (GeV)}
\psfrag{left}[]{\raisebox{0.75cm}{{\footnotesize 
$\Gamma_u^{(0)}(P_+<\Delta)$}}}
\psfrag{prelim}[]{{\footnotesize $^{\mu_i=1.5~{\rm GeV}}$}}
\psfrag{lo}[]{}
\psfrag{nlo}[]{}
\psfrag{nnlo}[]{}
\includegraphics[width=0.40\textwidth]{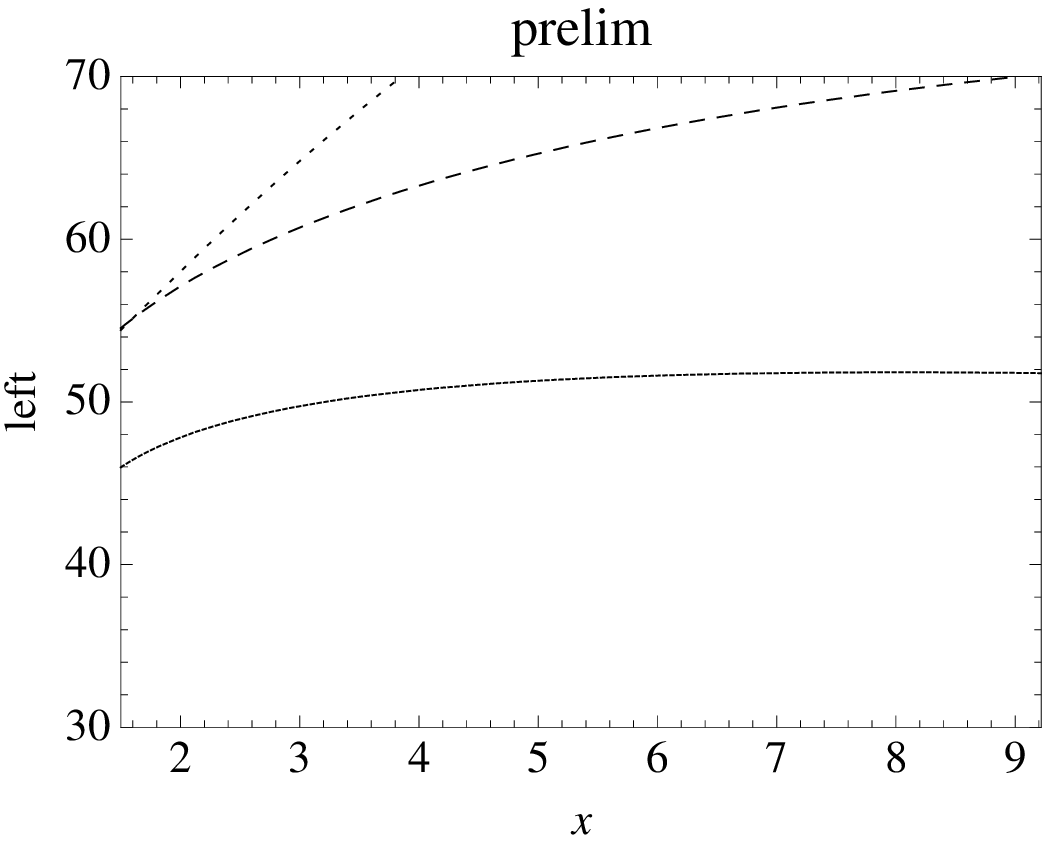} & 
\psfrag{x}[]{$\mu_i$ (GeV)}
\psfrag{left}[]{}
\psfrag{prelim}[]{{\footnotesize $^{\mu_h=m_b^*/\sqrt{2}}$}}
\psfrag{lo}[]{}
\psfrag{nlo}[]{}
\psfrag{nnlo}[]{}
\includegraphics[width=0.40\textwidth]{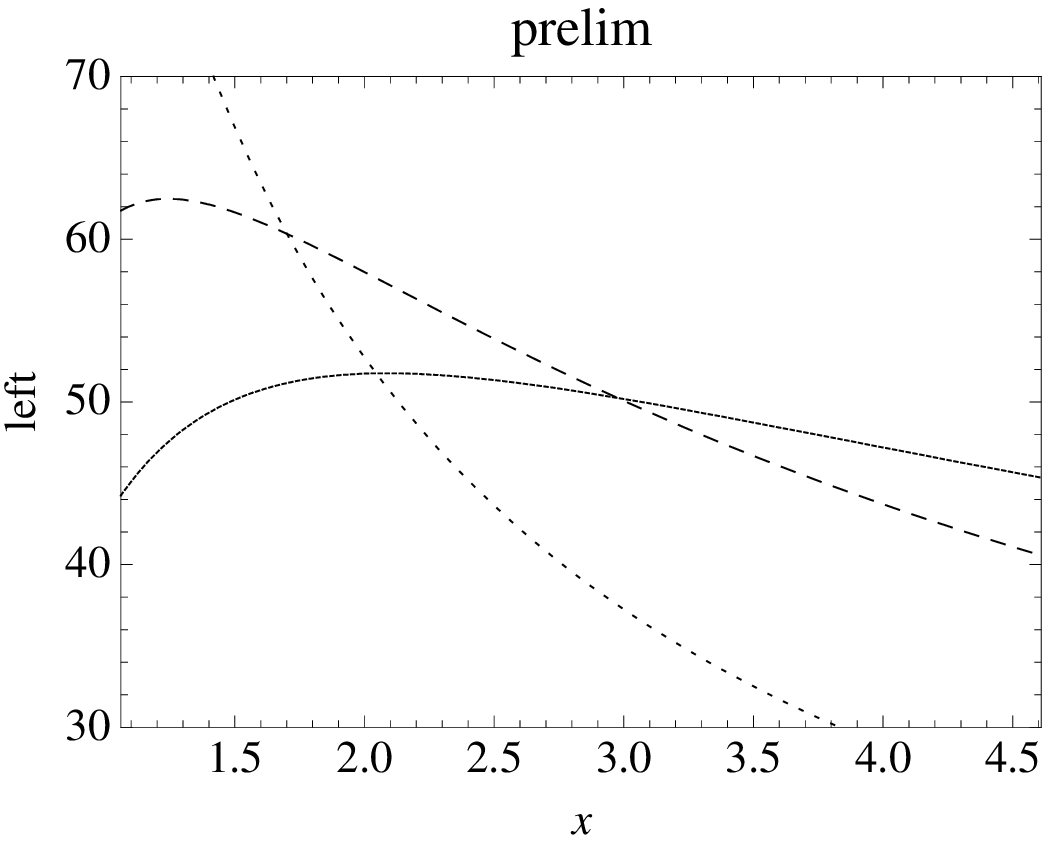} \\
& \\
&\\
\psfrag{x}[]{$\mu_h$ (GeV)}
\psfrag{left}[]{\raisebox{0.75cm} {\footnotesize 
$\Gamma_u^{(0)}(P_+<\Delta)$}}
\psfrag{prelim}[]{{\footnotesize $^{\mu_i=2.5~{\rm GeV}}$}}
\psfrag{lo}[]{}
\psfrag{nlo}[]{}
\psfrag{nnlo}[]{}
\includegraphics[width=0.40\textwidth]{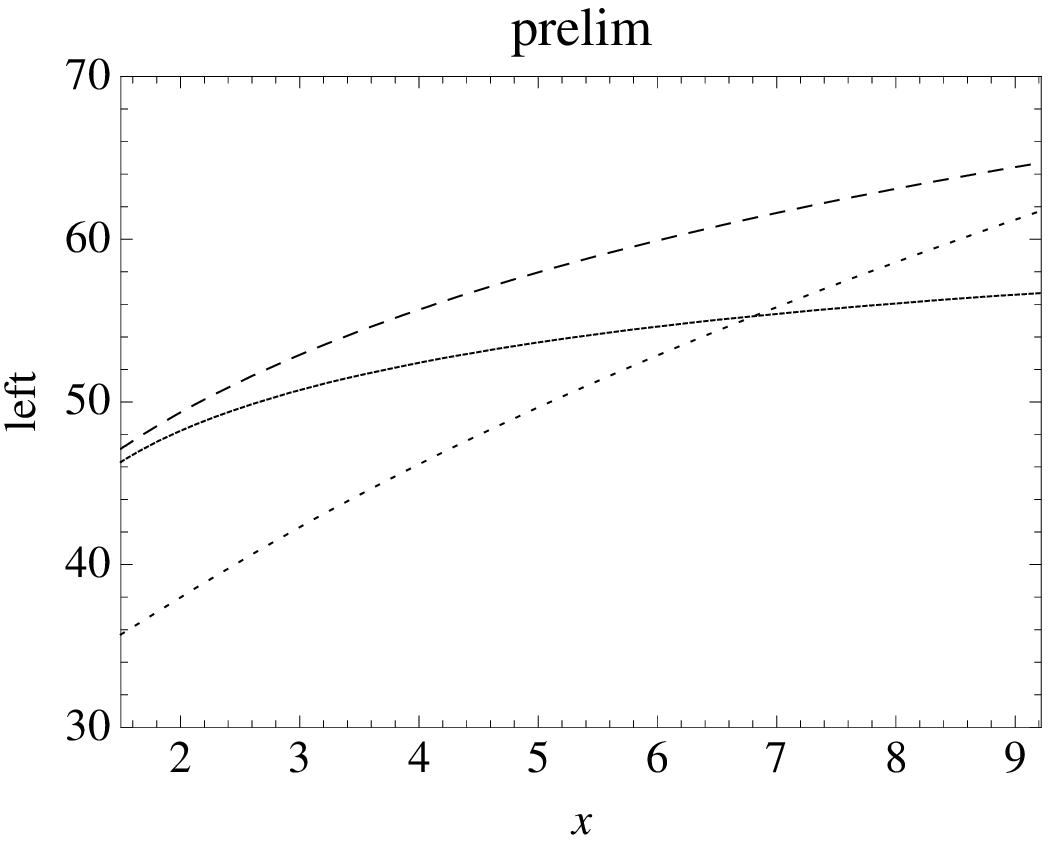} & 
\psfrag{x}[]{$\mu$ (GeV)}
\psfrag{left}[]{}
\psfrag{prelim}[]{{\footnotesize $^{\mu_h=\mu_i=\mu}$}}
\psfrag{FOPT}[]{{\small F.O.P.T.}}
\psfrag{nlo}[]{}
\psfrag{nnlo}[]{}
\includegraphics[width=0.40\textwidth]{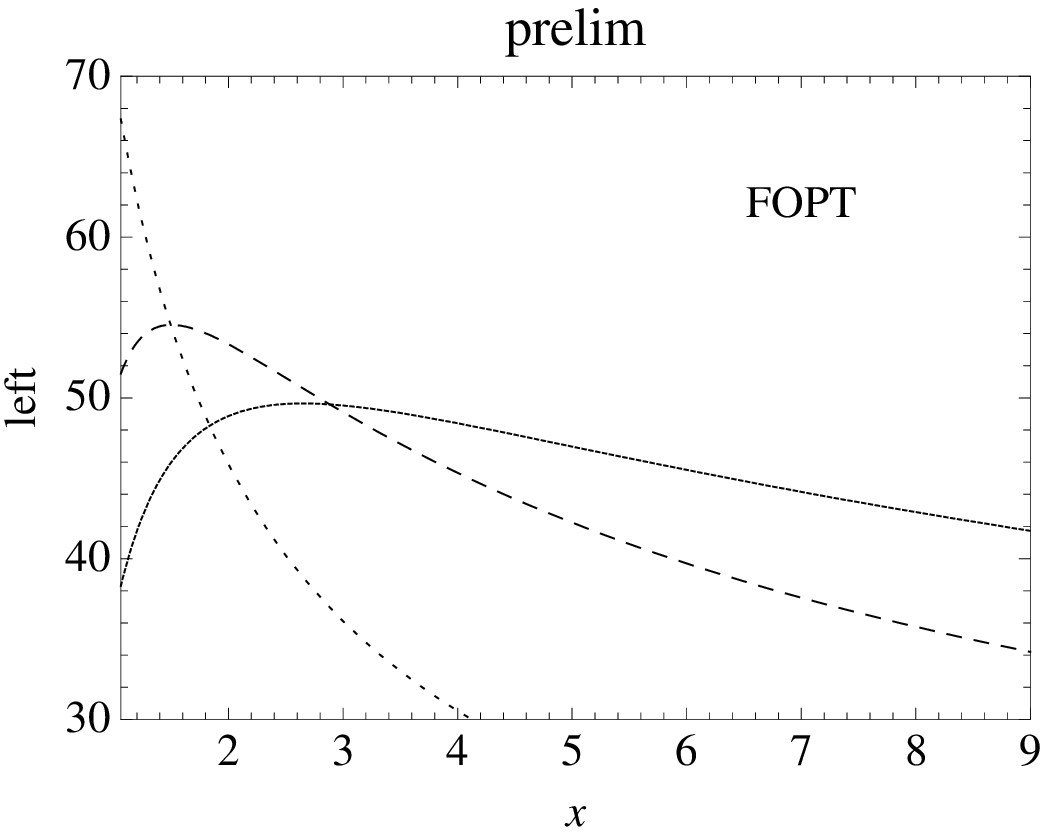}
\end{tabular}
\end{center}
\vspace{-0.5cm}
\caption{\label{fig:ScaleVar}  
Dependence of the partial 
rate $\Gamma_u^{(0)}$ (in units of $|V_{ub}|^2\,{\rm ps}^{-1}$) 
with a cut on $P_+<\Delta=0.66$~GeV on the matching scales
 $\mu_h$ and $\mu_i$ at LO (dotted), NLO (dashed), and NNLO (solid),
with the parameter choices $m_b^*=4.71$~GeV and $\mu_\pi^{*2}=0.2$~GeV$^2$.
The case $\mu_h=\mu_i=\mu$ corresponds to fixed-order perturbation
theory (F.O.P.T.). }
\end{figure}

\begin{figure}[t]
\begin{center}
\begin{tabular}{lr}
\hspace{-0.05\textwidth}
\psfrag{x}[]{$\mu_h$ (GeV)}
\psfrag{left}[]{ \raisebox{0.75cm}{{\footnotesize 
$\Gamma_u^{(0)}(M_X<M_0)$}}}
\psfrag{prelim}[]{{\footnotesize $^{\mu_i=1.5~{\rm GeV}}$}}
\psfrag{lo}[]{}
\psfrag{nlo}[]{}
\psfrag{nnlo}[]{}
\includegraphics[width=0.40\textwidth]{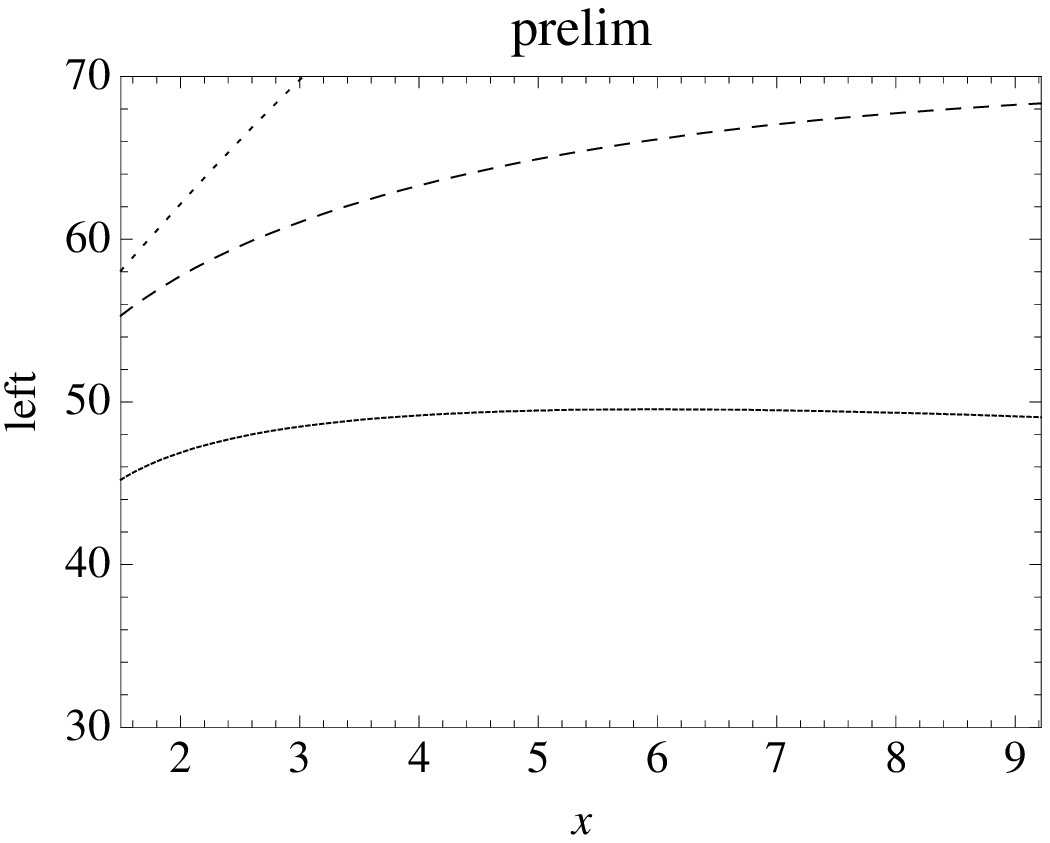} & 
\psfrag{x}[]{$\mu_i$ (GeV)}
\psfrag{left}[]{}
\psfrag{prelim}[]{{\footnotesize $^{\mu_h=m_b^*/\sqrt{2}}$}}
\psfrag{lo}[]{}
\psfrag{nlo}[]{}
\psfrag{nnlo}[]{}
\includegraphics[width=0.40\textwidth]{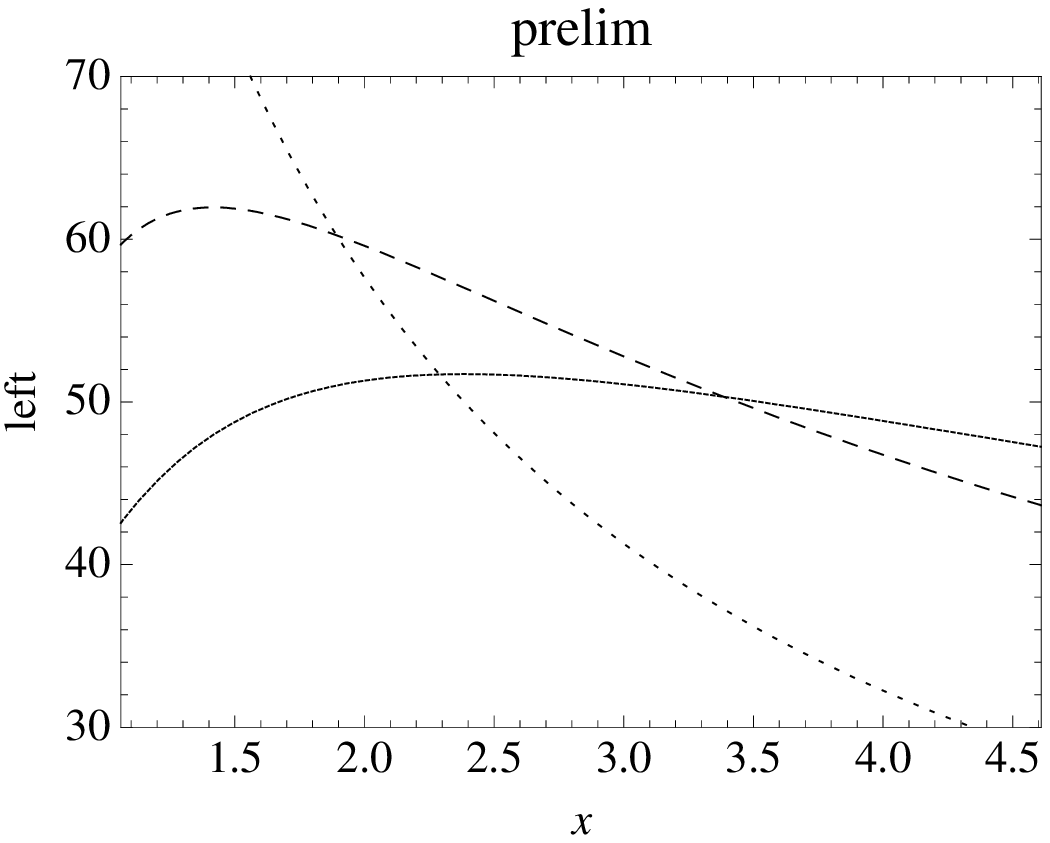} \\
& \\
& \\
\hspace{-0.05\textwidth}
\psfrag{x}[]{$\mu_h$ (GeV)}
\psfrag{left}[]{\raisebox{0.75cm}
{\footnotesize $\Gamma_u^{(0)}(M_X<M_0)$}}
\psfrag{prelim}[]{{\footnotesize $^{\mu_i=2.5~{\rm GeV}}$}}
\psfrag{lo}[]{}
\psfrag{nlo}[]{}
\psfrag{nnlo}[]{}
\includegraphics[width=0.40\textwidth]{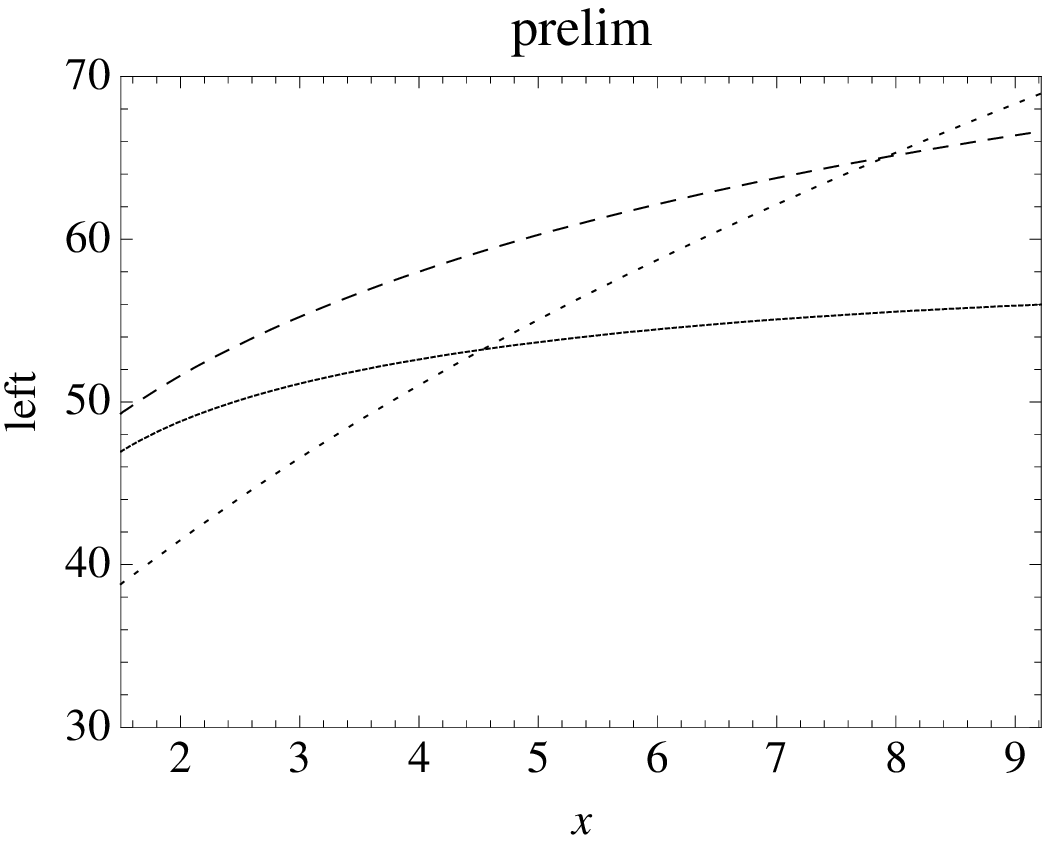} & 
\psfrag{x}[]{$\mu$ (GeV)}
\psfrag{left}[]{}
\psfrag{prelim}[]{{\footnotesize $^{\mu_h=\mu_i=\mu}$}}
\psfrag{FOPT}[]{{\small F.O.P.T.}}
\psfrag{nlo}[]{}
\psfrag{nnlo}[]{}
\includegraphics[width=0.40\textwidth]{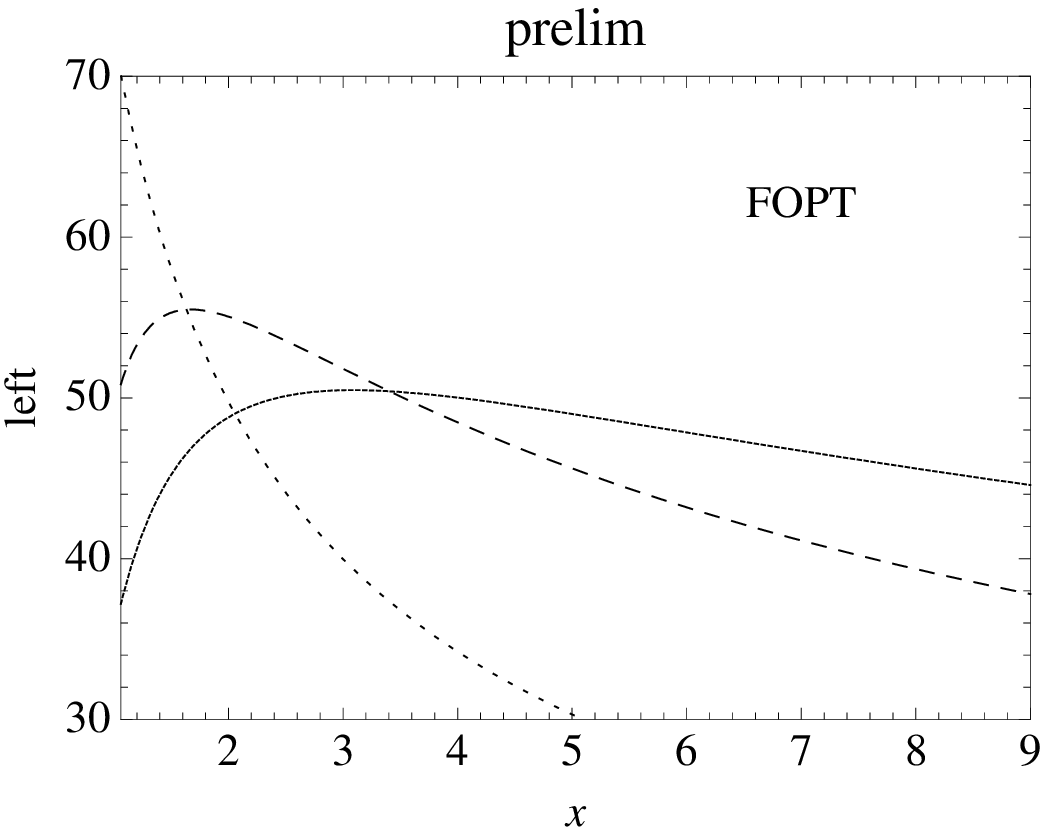}
\end{tabular}
\end{center}
\vspace{-0.5cm}
\caption{\label{fig:HadScaleVar}  
Dependence of the partial 
rate  $\Gamma_u^{(0)}$ (in units of $|V_{ub}|^2\,{\rm ps}^{-1}$)
with a cut on $M_X<M_0=1.7$~GeV on the matching scales
 $\mu_h$ and $\mu_i$ at LO (dotted), NLO (dashed), and NNLO (solid),
with the parameter choices $m_b^*=4.71$~GeV and $\mu_\pi^{*2}=0.2$~GeV$^2$.
The case $\mu_h=\mu_i=\mu$ corresponds to fixed-order perturbation
theory.   }
\end{figure}

\begin{figure}[t]
\begin{center}
\begin{tabular}{lr}
\hspace{-0.05\textwidth}
\psfrag{x}[]{$\mu_h$ (GeV)}
\psfrag{left}[]{\raisebox{0.75cm}{{\footnotesize 
$\Gamma_u^{(0)}(E_l>E_0)$}}}
\psfrag{prelim}[]{{\footnotesize $^{\mu_i=1.5~{\rm GeV}}$}}
\psfrag{lo}[]{}
\psfrag{nlo}[]{}
\psfrag{nnlo}[]{}
\includegraphics[width=0.40\textwidth]{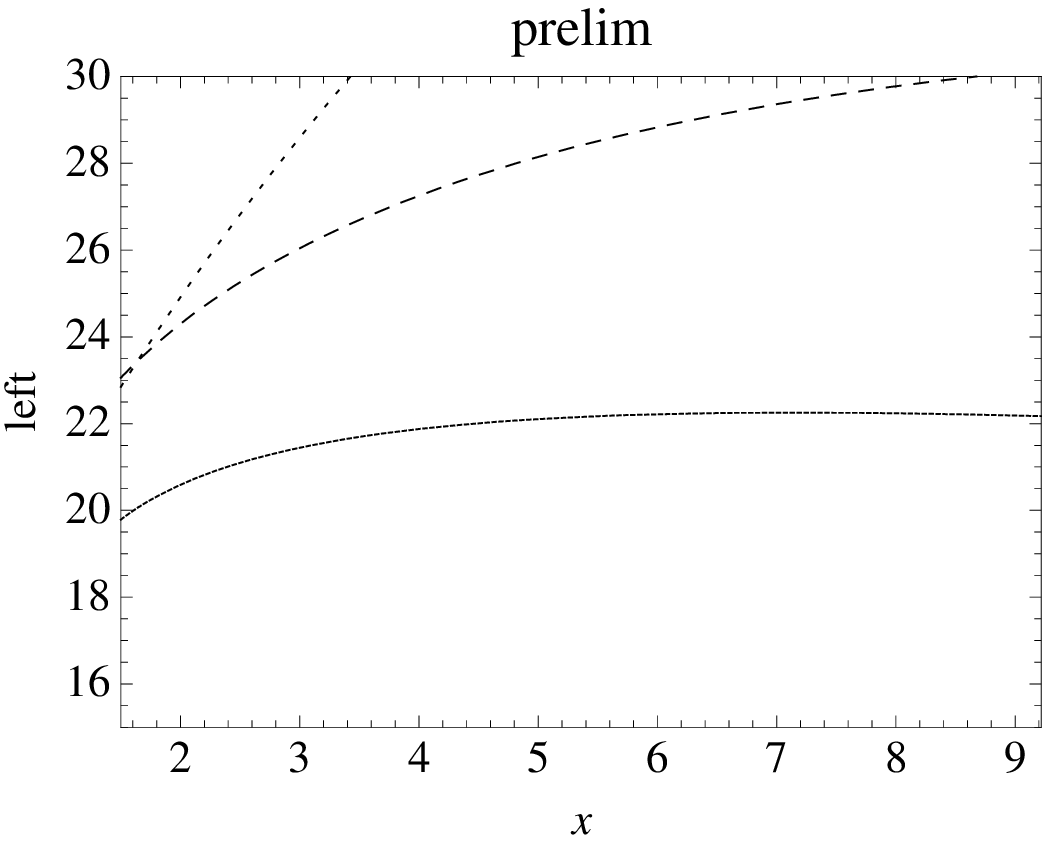} & 
\psfrag{x}[]{$\mu_i$ (GeV)}
\psfrag{left}[]{}
\psfrag{prelim}[]{{\footnotesize $^{\mu_h=m_b^*/\sqrt{2}}$}}
\psfrag{lo}[]{}
\psfrag{nlo}[]{}
\psfrag{nnlo}[]{}
\includegraphics[width=0.40\textwidth]{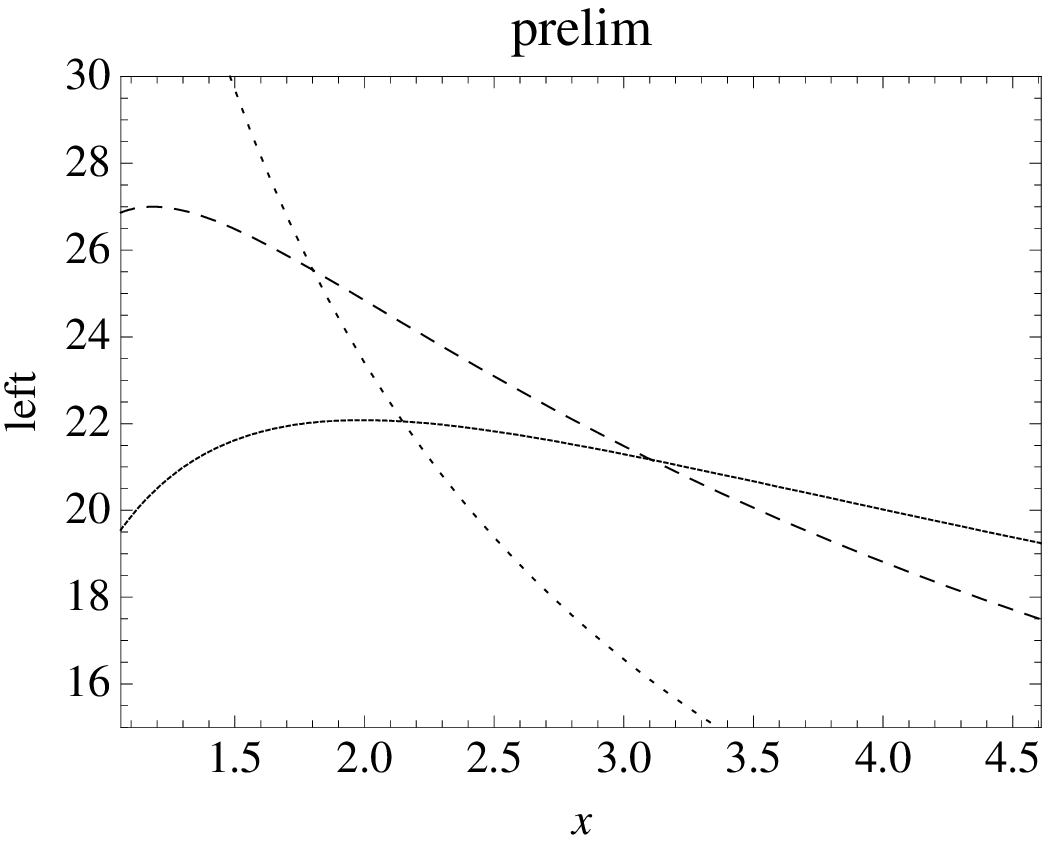}
\\
&\\
&\\
\hspace{-0.05\textwidth}
\psfrag{x}[]{$\mu_h$ (GeV)}
\psfrag{left}[]{\raisebox{0.75cm}{{\footnotesize 
$\Gamma_u^{(0)}(E_l>E_0)$}}}
\psfrag{prelim}[]{{\footnotesize $^{\mu_i=2.5~{\rm GeV}}$}}
\psfrag{lo}[]{}
\psfrag{nlo}[]{}
\psfrag{nnlo}[]{}
\includegraphics[width=0.40\textwidth]{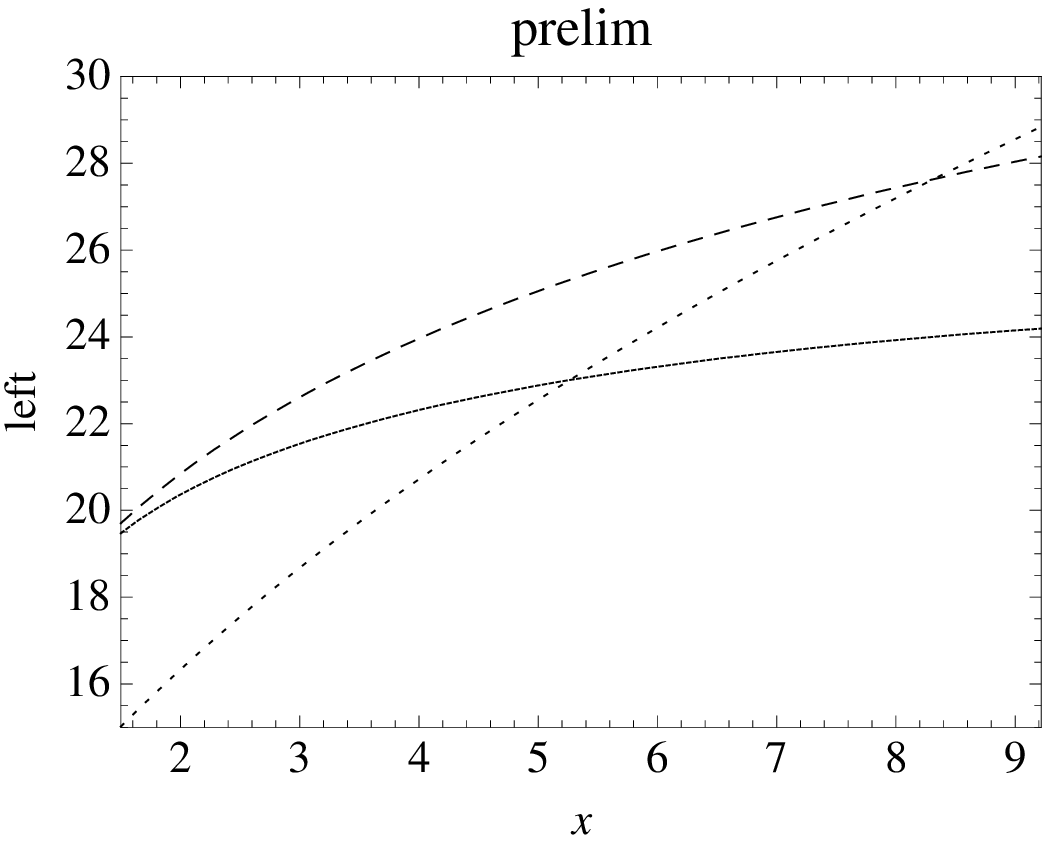} & 
\psfrag{x}[]{$\mu$ (GeV)}
\psfrag{left}[]{}
\psfrag{prelim}[]{{\footnotesize $^{\mu_h=\mu_i=\mu}$}}
\psfrag{FOPT}[]{{\small F.O.P.T.}}
\psfrag{nlo}[]{}
\psfrag{nnlo}[]{}
\includegraphics[width=0.40\textwidth]{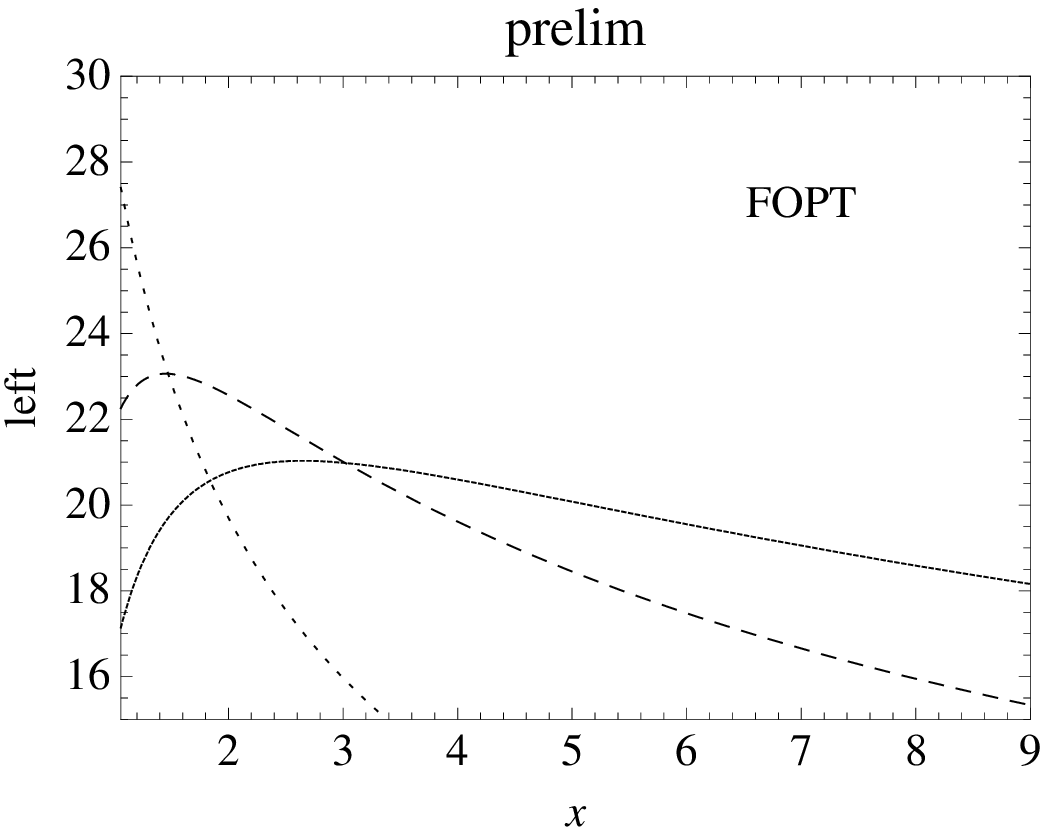}
\end{tabular}
\end{center}
\vspace{-0.5cm}
\caption{\label{fig:LepScaleVar}  
Dependence of the partial 
rate  $\Gamma_u^{(0)}$ (in units of $|V_{ub}|^2\,{\rm ps}^{-1}$) 
with a cut on $E_l>E_0=2.0$~GeV on the matching scales
$\mu_h$ and $\mu_i$ at LO (dotted), NLO (dashed), and NNLO (solid),
with the parameter choices $m_b^*=4.71$~GeV and $\mu_\pi^{*2}=0.2$~GeV$^2$.
The case $\mu_h=\mu_i=\mu$ corresponds to fixed-order perturbation
theory. }
\end{figure}
\begin{figure}[t]
\begin{center}
\begin{tabular}{lr}
\psfrag{x}[]{$\mu_i$ (GeV)}
\psfrag{left}[]{\raisebox{0.75cm}
{\footnotesize $\Gamma_u^{(0)}(P_+<\Delta)$}}
\psfrag{prelim}[]{{${\footnotesize^{\mu_h=\mu_i^2/(1.5~{\rm GeV})}}$}}
\psfrag{lo}[]{}
\psfrag{nlo}[]{}
\psfrag{nnlo}[]{}
\includegraphics[width=0.40\textwidth]{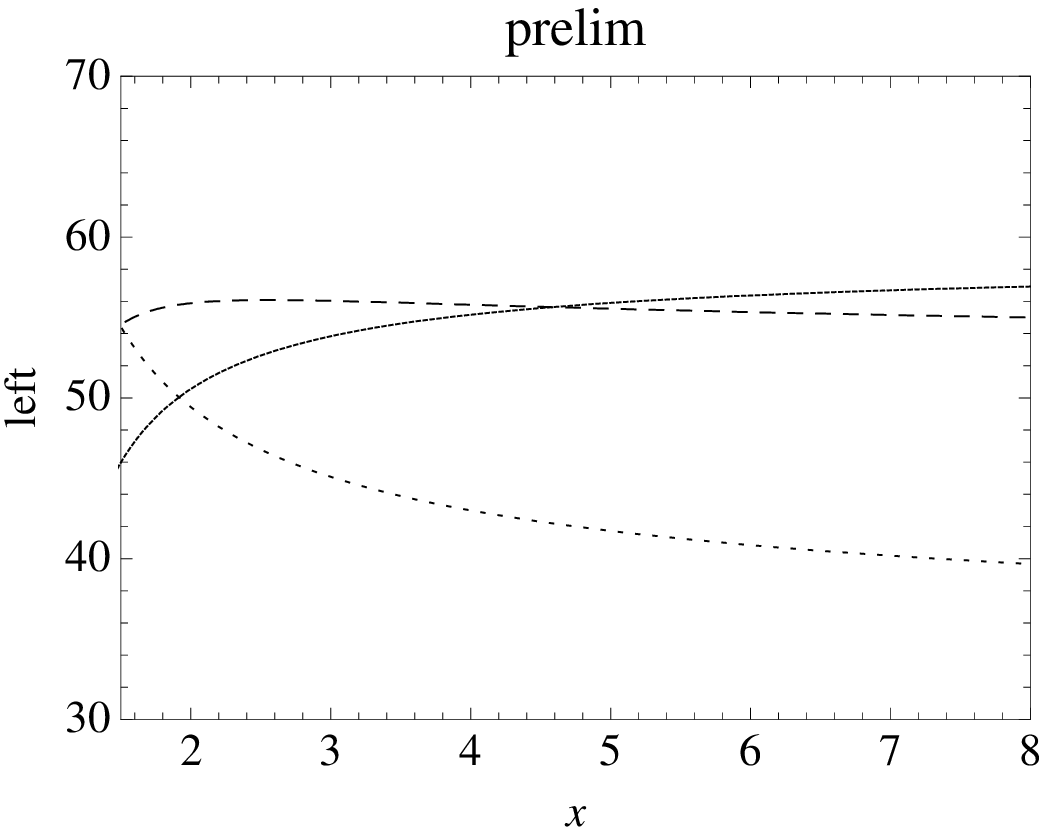}  & \hspace{1.2cm}
\psfrag{x}[]{$\mu_i$ (GeV)}
\psfrag{left}[]{\raisebox{0.75cm}
{{\footnotesize $\Gamma_u^{(0)}(M_X<M_0)$}}}
\psfrag{prelim}[]{{${\footnotesize^{\mu_h=\mu_i^2/(1.5~{\rm GeV})}}$}}
\psfrag{lo}[]{}
\psfrag{nlo}[]{}
\psfrag{nnlo}[]{}
\includegraphics[width=0.40\textwidth]{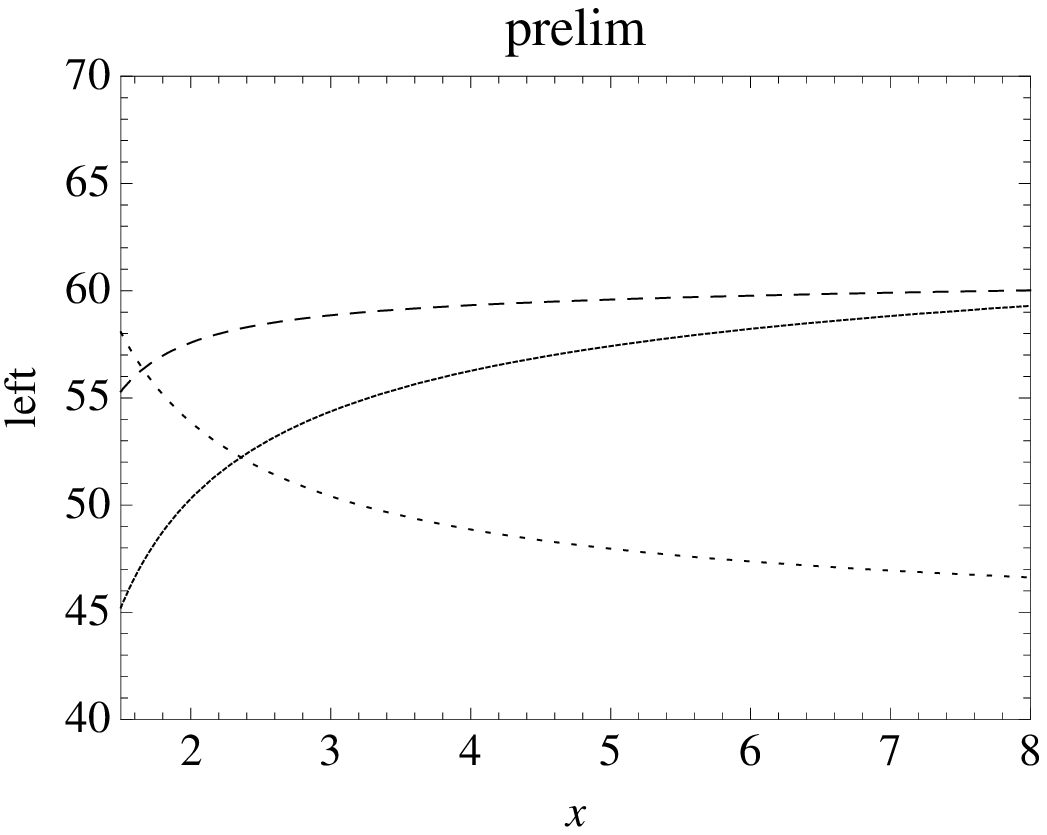} \\ & \\
\psfrag{x}[]{$\mu_i$ (GeV) }
\psfrag{left}[]{ \raisebox{0.75cm} 
{\footnotesize $\Gamma_u^{(0)}(E_l>E_0)$}}
\psfrag{prelim}[]{{${\footnotesize^{\mu_h=\mu_i^2/(1.5~{\rm GeV})}}$}}
\psfrag{lo}[]{}
\psfrag{nlo}[]{}
\psfrag{nnlo}[]{}
\includegraphics[width=0.40\textwidth]{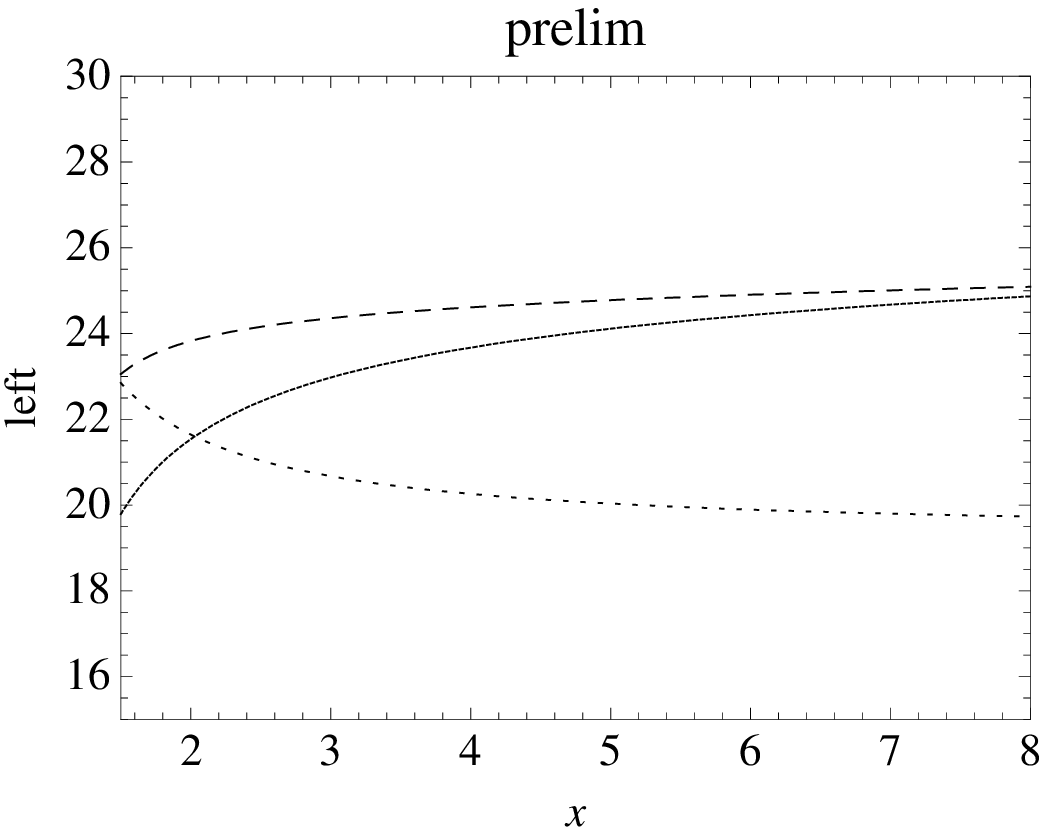} &
\end{tabular}
\end{center}
\vspace{-0.5cm}
\caption{\label{fig:MixedVar}  
Dependence of partial rates (in units of $|V_{ub}|^2\,{\rm ps}^{-1}$) 
on $\mu_i$, with $\mu_h=\mu_i^2/(1.5~{\rm GeV})$,  
at LO (dotted), NLO (dashed), and 
NNLO (solid), for the cuts $P_+<\Delta=0.66$~GeV, $M_X<M_0=1.7$~GeV,
and $E_l>E_0=2.0$~GeV. In each case $\mu_\pi^{*2}=0.2$~GeV$^2$. }
\end{figure}
\begin{figure}[t]
\begin{center}
\begin{tabular}{lr}
\psfrag{x}[]{$m_b^*$ (GeV)}
\psfrag{left}[]{\raisebox{0.75cm}
{\footnotesize $\Gamma_u^{(0)}(P_+<\Delta)$}}
\psfrag{prelim}[]{${\footnotesize^{\mu_h=4.25~{\rm GeV},\,\, \mu_i=2~{\rm GeV}}}$ }
\psfrag{lo}[]{}
\psfrag{nlo}[]{}
\psfrag{nnlo}[]{}
\includegraphics[width=0.40\textwidth]{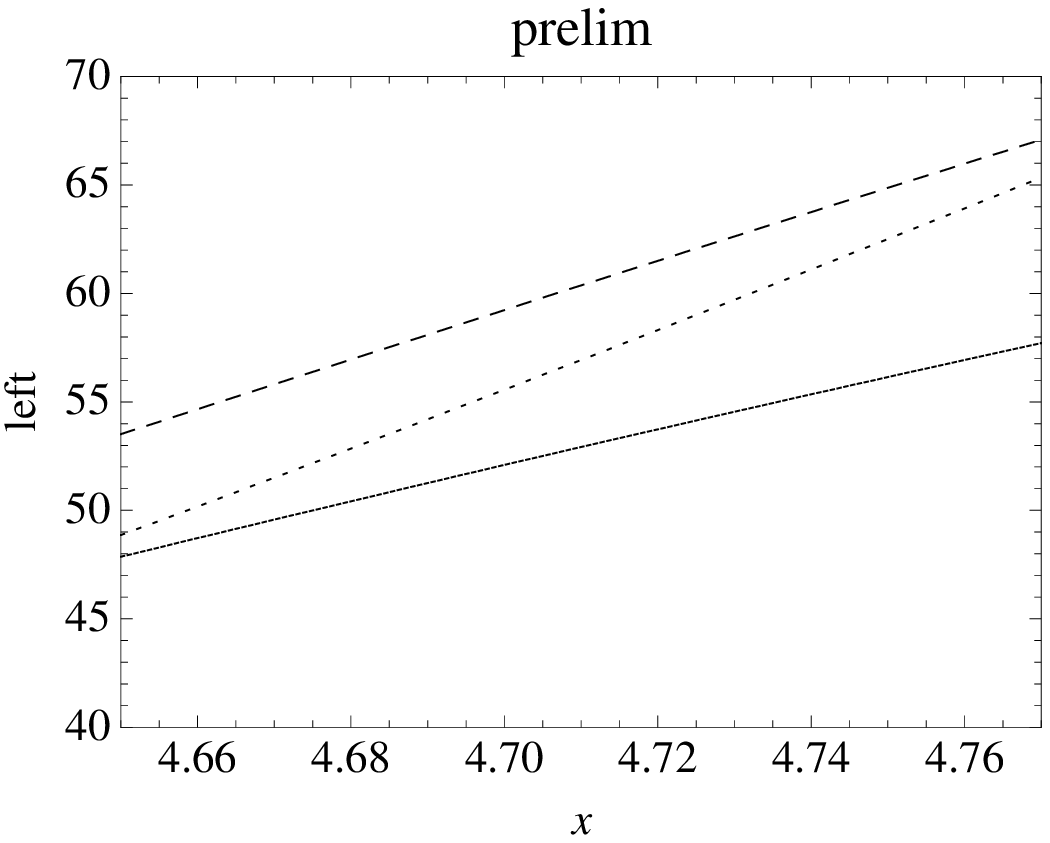}  & \hspace{1.2cm}
\psfrag{x}[]{$m_b^*$ (GeV)}
\psfrag{left}[]{\raisebox{0.75cm}
{{\footnotesize $\Gamma_u^{(0)}(M_X<M_0)$}}}
\psfrag{prelim}[]{{${\footnotesize^{\mu_h=4.25~{\rm GeV},\,\, \mu_i=2~{\rm GeV}}}$}}
\psfrag{lo}[]{}
\psfrag{nlo}[]{}
\psfrag{nnlo}[]{}
\includegraphics[width=0.40\textwidth]{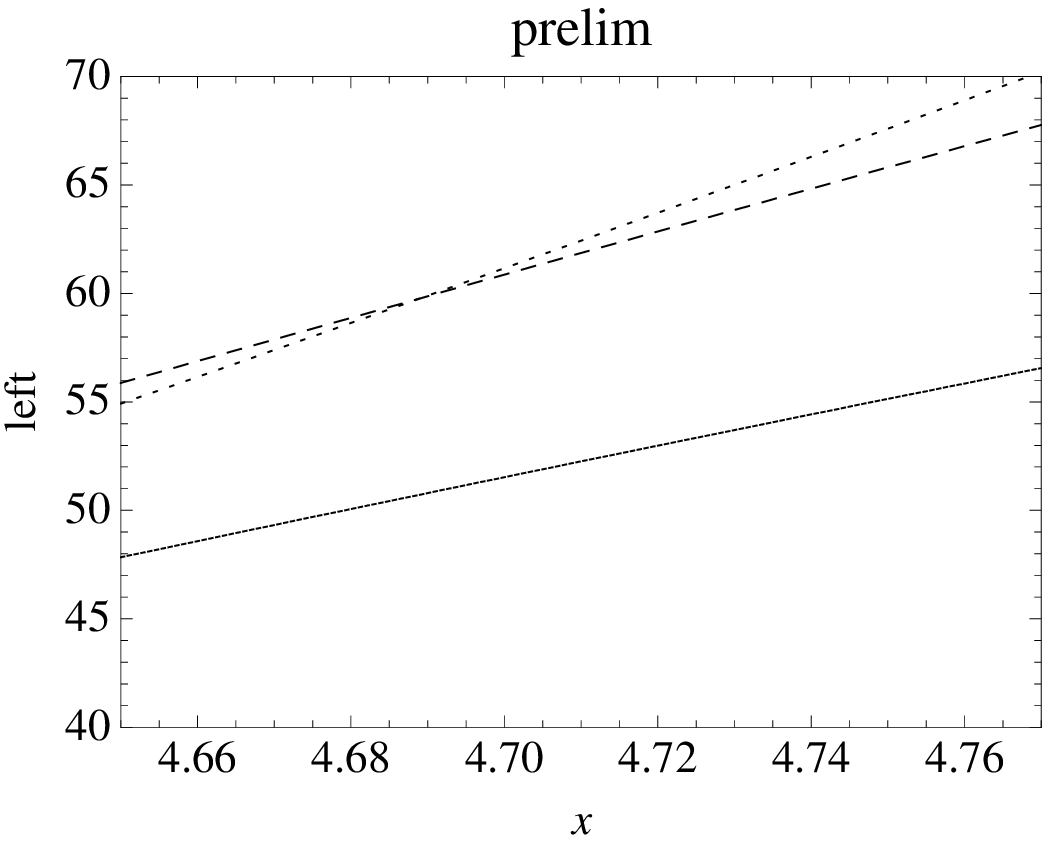} \\ & \\
\psfrag{x}[]{$m_b^*$ (GeV) }
\psfrag{left}[]{ \raisebox{0.75cm} 
{\footnotesize $\Gamma_u^{(0)}(E_l>E_0)$}}
\psfrag{prelim}[]{{${\footnotesize^{\mu_h=4.25~{\rm GeV},\,\, \mu_i=2~{\rm GeV}}}$}}
\psfrag{lo}[]{}
\psfrag{nlo}[]{}
\psfrag{nnlo}[]{}
\includegraphics[width=0.40\textwidth]{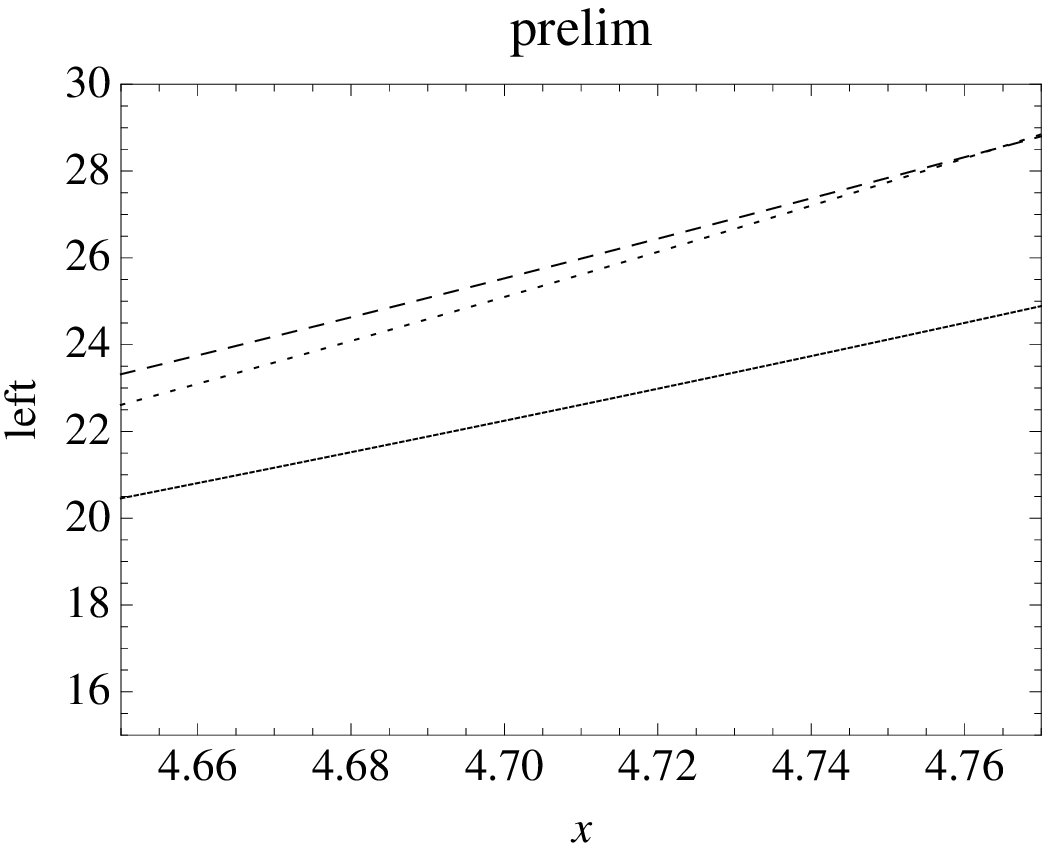} &
\end{tabular}
\end{center}
\vspace{-0.5cm}
\caption{\label{fig:mbVar}  
Dependence of partial rates (in units of $|V_{ub}|^2\,{\rm ps}^{-1}$) 
on $m_b^*$ at LO (dotted), NLO (dashed), and 
NNLO (solid), for the cuts $P_+<\Delta=0.66$~GeV, $M_X<M_0=1.7$~GeV,
and $E_l>E_0=2.0$~GeV. In each case $\mu_\pi^{*2}=0.2$~GeV$^2$. }
\end{figure}
\begin{figure}[t]
\begin{center}
\begin{tabular}{lr}
\psfrag{x}[]{$x$}
\psfrag{left}[]{\raisebox{0.75cm}
{{\footnotesize $\Gamma_u^{(0)}(P_+<\Delta)$}}}
\psfrag{prelim}[]{}
\psfrag{lo}[]{}
\psfrag{nlo}[]{}
\psfrag{nnlo}[]{}
\includegraphics[width=0.40\textwidth]{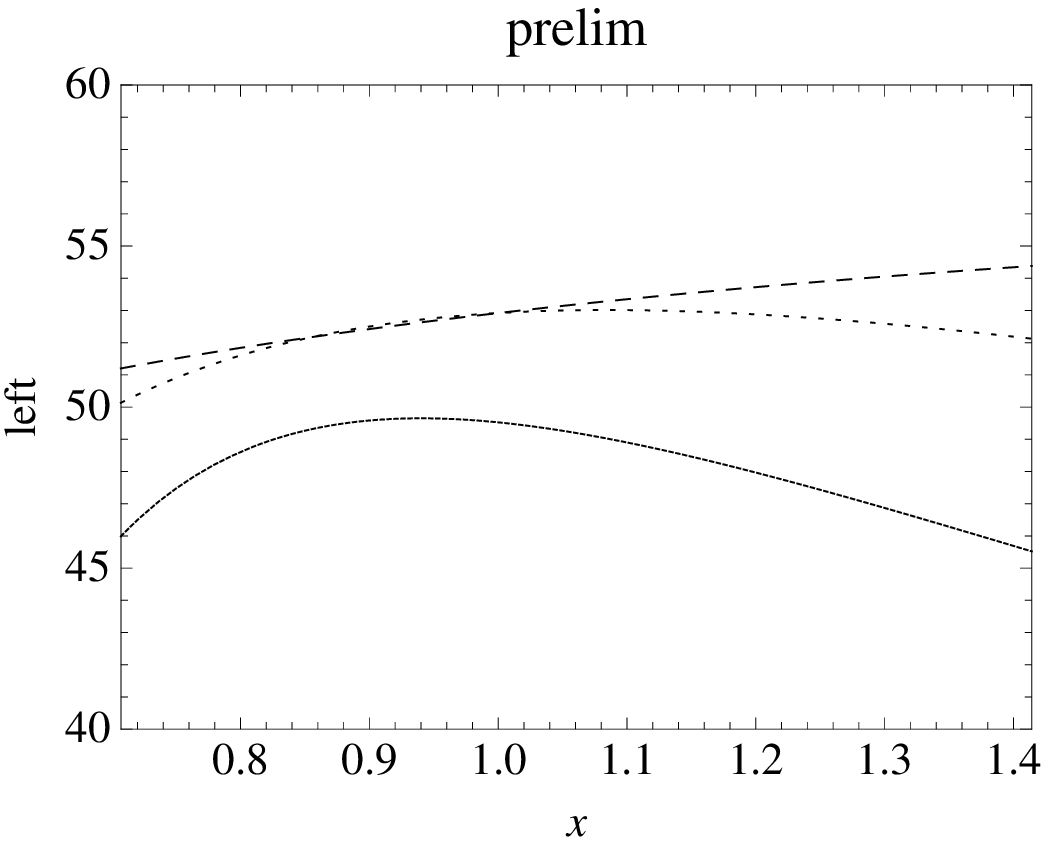}  & \hspace{1.2cm}
\psfrag{x}[]{$x$}
\psfrag{left}[]{\raisebox{0.75cm}
{{\footnotesize $\Gamma_u^{(0)}(M_X<M_0)$}}}
\psfrag{prelim}[]{}
\psfrag{lo}[]{}
\psfrag{nlo}[]{}
\psfrag{nnlo}[]{}
\includegraphics[width=0.40\textwidth]{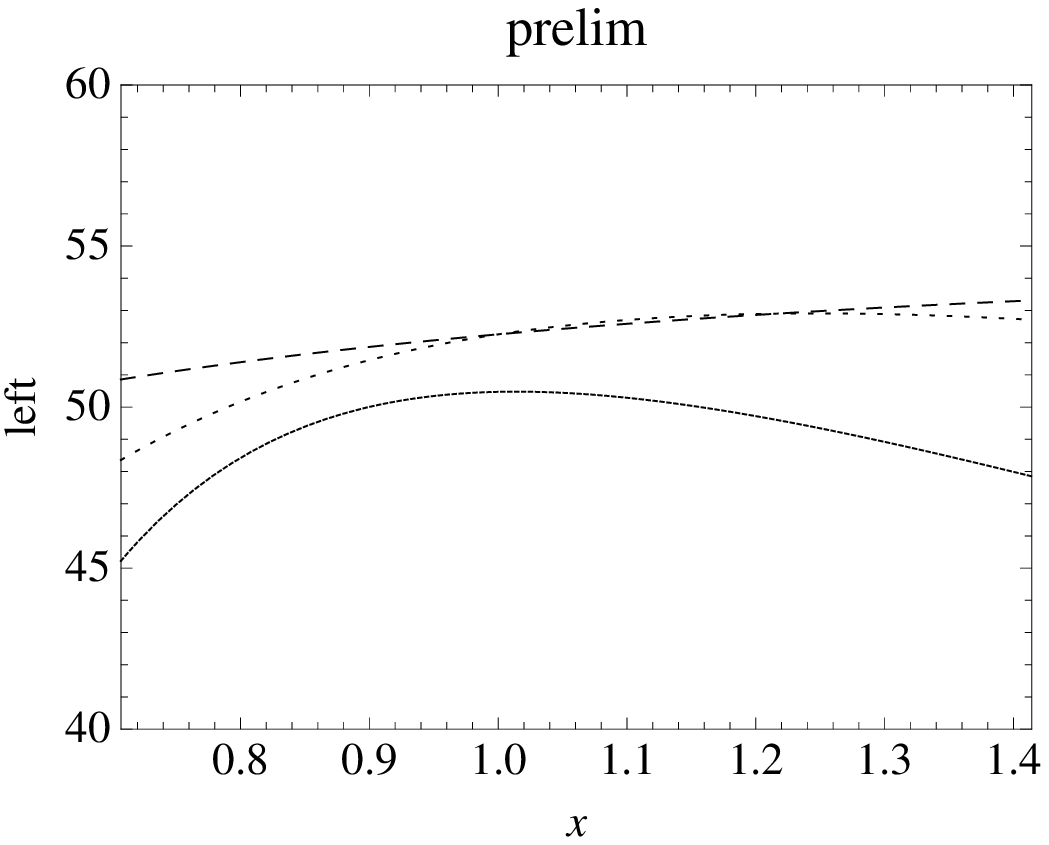} \\ & \\
\psfrag{x}[]{$x$ }
\psfrag{left}[]{\raisebox{0.75cm}{
{\footnotesize $\Gamma_u^{(0)}(E_l>E_0)$}}}
\psfrag{prelim}[]{{}}
\psfrag{lo}[]{}
\psfrag{nlo}[]{}
\psfrag{nnlo}[]{}
\includegraphics[width=0.40\textwidth]{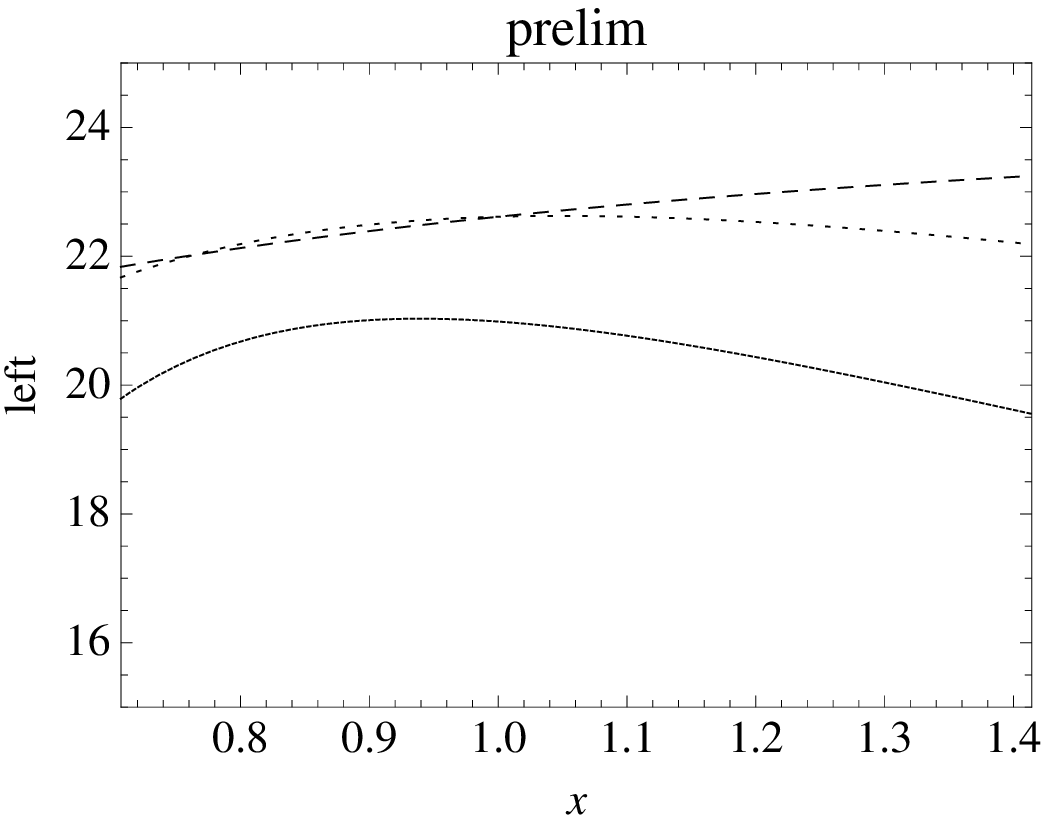} &
\end{tabular}
\end{center}
\vspace{-0.5cm}
\caption{\label{fig:FixedResummedScaleVar}  
Comparison of the NNLO partial rates  $\Gamma_u^{(0)}$ 
(in units of $|V_{ub}|^2\,{\rm ps}^{-1}$) in resummed perturbation
theory and fixed order, for the cuts
$P_+<\Delta=0.66$~GeV, $M_X<M_0=1.7$~GeV, and $E_l>E_0=2.0$~GeV.
The dashed line is the variation of 
the resummed results with $\mu_h= (4.25~{\rm GeV})x$, for
$\mu_i=2$~GeV, the dotted line is the variation with $\mu_i=(2~{\rm GeV})x$, 
for $\mu_h=4.25$~GeV, and the solid line is the variation of the fixed order result with $\mu=(3{\rm ~GeV})x^2$.  In all cases $m_b^*=4.71$~GeV and 
$\mu_\pi^{*2}=0.2$~GeV$^2$.}
\end{figure}

In this section we study the effects of  the NNLO perturbative corrections 
to the leading-power term $\Gamma_u^{(0)}$.  For 
the time being, we limit ourselves to three benchmark partial rates, 
defined by the following cuts:
\begin{enumerate}
\item   $P_+ < 0.66$~GeV  
\item  $M_X < 1.7$~GeV 
\item  $E_l > 2.0$~GeV 
\end{enumerate}
Experimental data is available for each of these partial decay
rates, and will be used to extract values of $|V_{ub}|$ in 
the next section. The goal of this section is to  
address the following questions:
\begin{itemize}
\item How does the NNLO analysis change the central values and error 
estimates compared to NLO?
\item Is resummation important, or is fixed-order perturbation 
theory sufficient?
\item How well do the NNLO corrections in the large-$\beta_0$ limit 
approximate the full results?
\end{itemize}
We shall examine these issues in the following subsections,
with the help of the information contained in Figures  
\ref{fig:ScaleVar}-\ref{fig:FixedResummedScaleVar}, and the 
 numbers given in Section \ref{subsec:BLM}.  Before interpreting the results,
we first make some comments on the numerical evaluations. 
To study the behavior of the perturbative expansion,  we give
results where  the structure functions (\ref{eq:Resummedf}) 
are evaluated at LO, NLO, and NNLO.  In doing this, we evaluate the product of 
matching functions $H\cdot \widetilde j$ and the RG-exponents
$S,\eta$, $a_{\gamma^{J}}$,  $a_{\gamma'}$ to the given order in $\alpha_s$.
These quantities also depend on the renormalization scheme for $m_b$, which
we choose as the shape-function scheme; the shifts from the pole
scheme are made using the perturbative relations (\ref{simplebeauty}) 
as appropriate at a given order in $\alpha_s$.
At this point we evaluate the resulting expressions numerically, 
without any further expansion.  This means, for example, 
that we do not use relations such as  
$y^{a_{\Gamma,{\rm LO}}+b\alpha_s}\approx y^{a_{\Gamma,{\rm LO}}}(1+b\,\alpha_s \ln y)$
(we have checked that the difference between the two expansions is negligible 
numerically).   Finally,  we use a common shape function when quoting
results at LO, NLO, and NNLO,  namely that where the moment 
constraints are implemented at NNLO.  For reference,
at $m_b^*=4.71$~GeV and
$\mu_{\pi}^{*2}=0.2$~GeV$^2$, the shape-function model is 
$\hat S(\hat \omega)=2.78~{\rm GeV}^{-1}\Omega^{1.29}\exp{(-\Omega) }$, where
$\Omega=\hat\omega\times 3.42~{\rm GeV}^{-1}$. 

A further issue is the treatment of the charm-quark mass.  The dependence
on this parameter first enters at NNLO, through diagrams where a charm loop is 
inserted into a gluon propagator.  
In evaluating the perturbative expressions, we always set $n_h=1$ and
work with $n_l=4$ light flavors. In other words, we treat $m_c\ll \mu_i$
and set $m_c=0$ when the fermion loops needed for the calculation of 
the two-loop hard and jet functions contain a charm quark.  
The calculations from  \cite{Bell:2008ws}
show that this is not a bad approximation 
to the full dependence in the hard function $H$.  Numerically,
it  may be more appropriate to treat $m_c\sim \mu_i$,
as discussed in the context of SCET in \cite{Boos:2005by, Boos:2005qx}, but 
a consistent treatment would also require the $m_c$ dependence in the 
jet function, which is not yet available.  

A final complication is that the analytic expression for the hard
function $H_{u1}$ contains a harmonic polylogarithm (HPL) of weight four, 
which cannot be expressed in terms of standard functions such 
as polylogarithms and their generalizations.  For the numerical
evaluation of HPLs we have used the \texttt{Mathematica} package
\texttt{HPL} \cite{Maitre:2005uu},  and the \texttt{FORTRAN} subroutine
\texttt{hplog} \cite{Gehrmann:2001pz}.

\subsection{Impact of NNLO corrections on central values and errors}
\label{subsec:NNLOrates}
The first question we wish to address is how the NNLO corrections
and implementation of the dependence on the scale $\mu_i$ affect
the central values and error estimates for the leading-power 
partial decay rates compared to the NLO analysis in \cite{Lange:2005yw}. 
We have shown in  Figures  \ref{fig:ScaleVar}-\ref{fig:LepScaleVar} a set
of plots meant to shed light on that issue.   To compare 
central values obtained  at different orders in perturbation
theory we have shown in the upper-left hand plots
of Figures  \ref{fig:ScaleVar}-\ref{fig:LepScaleVar} the dependence
on the scale $\mu_h$, for the fixed value $\mu_i=1.5$~GeV, which is
the BLNP choice.  We see that for this choice of $\mu_i$ 
the NNLO corrections are quite large in 
each case, and significantly lower the central values compared
to NLO.  At the same time, the perturbative uncertainties associated 
with the matching scale $\mu_h$ are reduced, as indicated by 
the flattening of the curves at higher orders in perturbation theory.

A new element of our analysis compared to the procedure in 
\cite{Lange:2005yw} is that the matching scale $\mu_i$ is not 
necessarily fixed to the value $\mu_i=1.5$~GeV.  In the upper right-hand
plots of  Figures  \ref{fig:ScaleVar}-\ref{fig:LepScaleVar}, we 
display the dependence of the results on the scale $\mu_i$, for 
fixed values $\mu_h=m_b^*/\sqrt{2}$.  We note a large dependence
on the intermediate scale at NLO, which is reduced but still 
significant at  NNLO.  Because the dependence on $\mu_i$ is so 
strong at NLO, even small changes of the value of $\mu_i$ can
can alter rather drastically the agreement between the NLO and NNLO results.  
To illustrate this effect we have shown in the bottom left-hand
plots of Figures  \ref{fig:ScaleVar}-\ref{fig:LepScaleVar} the 
dependence of the partial rates on $\mu_h$ for $\mu_i=2.5$~GeV,
which is higher than the BLNP choice.  This higher value
of the intermediate scale brings the NLO results into closer agreement
with the NNLO results.

 So far, we have fixed one of the two matching scales to a default
value and studied the behavior of the partial decay rates under variations
of the other. Another option is to establish a correlation between the scales 
and vary them simultaneously.  Setting them equal to one another corresponds 
to fixed-order perturbation theory and will be discussed in the next 
section.  A different choice, which is somewhat more natural in SCET, 
is to vary the scales such that the relation $\mu_h\sim \mu_i^2/\mu_s$ is 
respected, where $\mu_s$ is a typical soft scale.  
In Figure \ref{fig:MixedVar}, we show the behavior of the partial 
rates under such a correlated variation for the choice $\mu_s=1.5$~GeV.
It is remarkable that the NLO result is almost entirely stable under 
scale variations.  This feature appears to be accidental, since 
at NNLO the scale dependence actually becomes stronger under this
correlated running.

In addition to the unphysical dependence on the matching 
scales $\mu_h$ and $\mu_i$,  the partial rates contain a rather
strong parametric dependence on the numerical value of  
the $b$-quark mass $m_b^*,$ mainly 
through the moment constraints 
on the shape-function model; scaling relations for this dependence
were derived in \cite{Lange:2005yw}.  We have already shown in 
Figure~\ref{fig:SFs} how changing $m_b^*$ distorts the shape-function 
model. We show in  Figure~\ref{fig:mbVar} how these changes in
the model translate into changes in the partial rates in the range 
$4.65 \,{\rm GeV} < m_b^* < 4.77$~GeV. Obviously, the present 
uncertainty in the value of the  $b$-quark mass adds a 
large parametric uncertainty to the analysis.
 
From the above discussion we can conclude that, compared to 
BLNP analysis in \cite{Lange:2005yw}, where $\mu_i=1.5$~GeV,
the net effect of the NNLO corrections is to lower the 
central values for partial 
rates by around $15$--$20\%$, while at the same time reducing 
the perturbative uncertainty on the scale $\mu_h$.  Moreover, 
we have pointed out that there is a considerable dependence
on the scale $\mu_i$ at NLO, which is reduced but still
significant even at NNLO. 
To make these statements more quantitative, we give central 
values and error estimates for some input parameters.  
In particular, we choose the  values $\mu_{i}^{\rm def}=2.0$~GeV 
and  $\mu_{h}^{\rm def}=4.25$~GeV, and then vary the scales in the range 
$1/\sqrt{2}< \mu_{h,i}/\mu_{h,i}^{\rm def}<\sqrt{2}$.  
We then find for the three benchmark partial rates, in units 
of $|V_{ub}|^2 \, {\rm ps}^{-1}$:
\begin{itemize}

\item $P_+<0.66$~GeV:

\vspace{0.4cm}
\begin{tabular}{|c|c|c|c|}
\hline
 & $\Gamma_u^{(0)}$ & $\mu_h$ & $\mu_i$   \\ \hline \hline
NLO & 60.37 & {\footnotesize $^{+3.52}_{-3.37}$} & {\footnotesize $^{+3.81}_{-6.67}$} 
\\
NNLO & 52.92  & {\footnotesize $^{+1.46}_{-1.72}$} & {\footnotesize $^{+0.09}_{-2.79}$}\\
\hline 
\end{tabular}

\item $M_X<1.7$~GeV:

\vspace{0.4cm}
\begin{tabular}{|c|c|c|c|}
\hline
 & $\Gamma_u^{(0)}$ & $\mu_h$ & $\mu_i$  \\ \hline \hline
NLO & 61.86 & {\footnotesize $^{+3.23}_{-3.21}$} & {\footnotesize $^{+1.89}_{-5.50}$} 
\\
NNLO & 52.26  & {\footnotesize $^{+1.05}_{-1.40}$} & {\footnotesize $^{+0.65}_{-3.90}$}\\
\hline 
\end{tabular}

\item $E_l>2.0$~GeV:

\vspace{0.4cm}
\begin{tabular}{|c|c|c|c|}
\hline
 & $\Gamma_u^{(0)}$ & $\mu_h$ & $\mu_i$  \\ \hline \hline
NLO & 25.98 & {\footnotesize $^{+1.63}_{-1.61}$} & {\footnotesize $^{+1.69}_{-2.81}$} 
\\
NNLO & 22.61  & {\footnotesize $^{+0.63}_{-0.78}$} & {\footnotesize $^{+0.01}_{-0.94}$}\\
\hline 
\end{tabular}

\end{itemize}
The default numbers refer to the value of the partial rate for 
$\mu_{h,i}=\mu_{h,i}^{\rm def}$.  To obtain the uncertainties associated
with  variations of $\mu_h$, we keep $\mu_i$ fixed to its default 
value, and then assign upper (lower) errors by 
picking out the highest (lowest) value of the decay rate in 
the range $1/\sqrt{2}<\mu_h/\mu_h^{\rm def}<\sqrt{2}$; the procedure
for $\mu_i$ is analogous.

\subsection{Comparison with fixed-order perturbation theory}
In this section we compare the results from
RG-improved perturbation theory with those obtained in a fixed-order
calculation.   The standard argument to motivate the use of 
resummed perturbation theory is that even though logs of $\mu_h/\mu_i$ 
are not particularly large, the running coupling constant runs 
quickly at low scales  $\mu_i\sim 1.5$~GeV, so the 
scale separation between the hard and
jet scales is important. If this were the case, we would expect 
to see a large scale dependence in the fixed-order result.
To test whether this actually happens, we have shown in 
the bottom right plots of Figures  \ref{fig:ScaleVar}-\ref{fig:LepScaleVar}
the results where the matching scales are set to $\mu_h=\mu_i=\mu$, which
corresponds to fixed-order perturbation theory.  
We see that the NNLO results are rather stable under variations of $\mu$
from 1.5~GeV to $2m_b^*$. Also, the convergence of the perturbation
series is actually better than in resummed perturbation theory, in the 
sense that the NNLO results lie within the ranges covered by 
varying the renormalization scale $\mu$ in the LO and NLO results. 
   
To illustrate further  the differences between resummed and fixed-order 
perturbation theory we show in 
Figure~\ref{fig:FixedResummedScaleVar} the NNLO results 
for both cases.  After the substitutions indicated in 
the caption, the curves can be used to compare 
the fixed-order results in the range
$\mu=\left(1.5-6\right)$~GeV with the resummed results in the ranges 
$1/\sqrt{2}< \mu_{h,i}/\mu_{h,i}^{\rm def}<\sqrt{2}$, with 
$\mu_{i}^{\rm def}=2.0$~GeV and  $\mu_{h}^{\rm def}=4.25$~GeV. 
In this way, the scale in the fixed-order result is varied
in a range which very nearly covers the 
lowest value of $\mu_i$ to the highest value of $\mu_h$.  
To compare results more closely,  we list tables of 
central values and errors, to be compared with the resummed 
results in the previous section.  We choose the default
scale as $\mu=3.0$~GeV, and quote the uncertainty obtained by 
varying it up and down by a factor of two.  We then
find, in units of $|V_{ub}|^2\,{\rm ps^{-1}}$:

\begin{itemize}

\item $P_+<0.66$~GeV:

\vspace{0.3cm}

\begin{tabular}{|c|c|c|}
\hline
Fixed-Order & $\Gamma_u^{(0)}$ & $\mu$  \\ \hline \hline
NLO & 49.11 & {\footnotesize $^{+5.43}_{-9.41}$}  \\
NNLO & 49.53 & {\footnotesize $^{+0.13}_{-4.01}$} \\
\hline 
\end{tabular}

\item $M_X<1.7$~GeV:

\vspace{0.4cm}

\begin{tabular}{|c|c|c|}
\hline
Fixed-Order &  $\Gamma_u^{(0)}$ & $\mu$    \\ \hline \hline
NLO & 51.81 & {\footnotesize $^{+3.69}_{-8.62}$}  
\\
NNLO & 50.47 & {\footnotesize $^{+0.01}_{-2.62}$} 
  \\
\hline 
\end{tabular}

\item $E_l>2.0$~GeV:

\vspace{0.4cm}

\begin{tabular}{|c|c|c|}
\hline
Fixed-Order & $\Gamma_u^{(0)}$ & $\mu$     \\ \hline \hline
NLO & 21.01 & {\footnotesize $^{+2.04}_{-3.54}$} 
\\
NNLO & 20.99 & {\footnotesize $^{+0.04}_{-1.43}$}\\
\hline 
\end{tabular}

\end{itemize}

\vspace{0.4cm}

\noindent At NNLO the errors are comparable with the resummed
results from the previous section. In each case the 
central values for the fixed-order results are noticeably lower 
than the resummed ones. 

The analysis above indicates that, for the leading-order term in the 
SCET expansion, the factorization of the perturbative coefficient
multiplying the leading-order shape function into jet and hard
functions is not strictly necessary: using
the fixed-order results does not lead to large scale uncertainties
compared to the treatment in RG-improved perturbation theory, nor 
to a poor convergence of the perturbative expansion.
Given this fact, it is legitimate to ask whether ``kinematic'' power
corrections, namely those suppressed by the perturbative ratio
of jet to hard scales and scaling as $P_+/P_-\sim \Lambda_{\rm QCD}/m_b$
in the shape-function region, should be treated separately from 
the leading-order term.  The one-loop analysis in  \cite{Lange:2005yw}
shows that the $1/m_b$ expansion of these kinematic power corrections
converges very quickly, and that the leading-order term is indeed of
the size expected of a $1/m_b$ correction.
To check this at two loops would require the calculation of the 
hadronic tensor in fixed-order perturbation theory to this order, 
which has yet to be done.  While partial results in the 
large-$\beta_0$ limit are available,  we shall see in the 
next section that, for the leading-order term, they do not accurately 
approximate the full two-loop calculations. 
In absence of evidence to the contrary, we shall assume
that ignoring the kinematically suppressed two-loop corrections provides 
an accurate approximation of decay rates to this order.

\subsection{NNLO corrections in the large-${\bm \beta_0}$ limit}
\label{subsec:BLM}

We now give illustrative results for the case where
the large-$\beta_0$ approximation to the two-loop hard and 
jet functions is used.  To obtain these results, we
pick out the  $n_lC_F$ color structure in the $\alpha_s^2$ corrections to
$H\cdot \widetilde j$, and then set $n_l\to -3 \beta_0/2$. 
As a check on our results, we have confirmed that convoluting
the resulting expression with the large-$\beta_0$ limit of the
partonic shape function evaluated
at NNLO in \cite{Becher:2005pd}, and setting all matching scales
to a common scale $\mu$, we reproduce the NNLO corrections to the 
partial rate with a cut on $P_+$ obtained in the large-$\beta_0$ limit
in \cite{Campanario:2008xu} (these can also be obtained from
the triple differential result in \cite{Aquila:2005hq}). 
To compare numerical values of partial decay
rates in the large-$\beta_0$ limit with the exact ones,  we evaluate 
the hard and jet  functions at a common scale $\mu_h=\mu_i=1.5$ GeV.  
Furthermore,  we choose $m_b^*=4.71$~GeV and $\mu_\pi^{*2}=0.2$~GeV$^2$,
and convolute both results with a common shape function obtained by 
implementing the moment constraints at NNLO. 
Then the full NNLO results for $\Gamma_u^{(0)}$, labeled
``QCD'', and the large-$\beta_0$ 
results, labeled ``BLM'', read for the three cuts (in units 
of $|V_{ub}^2|\,{\rm ps^{-1}}$):
\begin{itemize}
\item $P_+<0.66$~GeV:
{\small \begin{eqnarray}
&&\mbox{QCD:} \, \, \,  54.43+
0.11 \,[\alpha_s]
+ \big(-3.68\, [h] - 0.26 \,[j] 
-4.61 \,[ h\, j] =-8.55 \big)\,[\alpha_s^2]=45.99 \nonumber \\
&&\mbox{BLM:} \, \, \,  54.43+
0.11 \,[\alpha_s]
+ \big(-14.1\, [h] + 14.1 \,[j] 
\hspace{1.92cm} =- 0.02 \big)\,[\alpha_s^2] =54.52 \nonumber
\end{eqnarray}}
\item $M_X<1.7$~GeV:
{\small  \begin{eqnarray}
&&\mbox{QCD:} \, \, \,  58.06-
2.76 \,[\alpha_s]
+ \big(-4.05\, [h] - 1.99 \,[j] 
-4.04 \,[ h\, j] =-10.1 \big)\,[\alpha_s^2]=45.22 \nonumber \\
&&\mbox{BLM:} \, \, \,  58.06-
2.76 \,[\alpha_s]
+ \big(-15.7\, [h] + 13.0 \,[j] 
\hspace{1.92cm} =-2.68 \big)\,[\alpha_s^2] = 52.62 \nonumber
\end{eqnarray}}
\item $E_l>2.0$~GeV:
{\small  \begin{eqnarray}
&&\mbox{QCD:} \, \, \,  22.85+
0.21 \,[\alpha_s]
+ \big(-1.64\, [h] + 0.65 \,[j] 
-2.28 \,[ h\, j] =-3.27 \big)\,[\alpha_s^2]=19.79 \nonumber \\
&&\mbox{BLM:} \, \, \,  22.85+
0.21 \,[\alpha_s]
+ \big(-7.12\, [h] + 6.32 \,[j] 
\hspace{1.92cm} =-0.80 \big)\,[\alpha_s^2] = 22.26 \nonumber
\end{eqnarray}}
\end{itemize}
We have included labels to distinguish the contributions 
from different orders in $\alpha_s$, and for the 
NNLO pieces we  have also indicated  whether the contribution 
comes from  the hard function ($[h])$, 
the jet function ($[j]$), or the product of one-loop hard and jet 
functions ($[h\,j]$). 
We see that the contributions
of the two-loop hard and jet functions in the large-$\beta_0$
approximation are larger than the full results. 
In each case, they undergo large cancellations in the sum and
give results which are much smaller in magnitude than the complete 
ones.  Given the poor agreement of results in the
large-$\beta_0$ approximation with the full NNLO results for the 
leading order term in the HQET expansion, it is 
somewhat questionable that including only such terms leads to 
any phenomenological improvement compared to NLO. 
For the jet function $\widetilde j$, the poor agreement between 
the full results and  those in the large-$\beta_0$ limit was previously noted 
in \cite{Becher:2006qw}.

\section{Impact on the extraction of  $\bm{|V_{ub}|}$}
\label{sec:Vub}
In this section we study the numerical impact of our results 
on the extraction of $|V_{ub}|$ from experimental information 
on partial decay rates.  In order to do so, we combine our 
NNLO results for the leading-power term $\Gamma_u^{(0)}$ 
with the power corrections obtained in  \cite{Lange:2005yw}.
A given partial rate $\Gamma_u$ is then obtained in the form
\begin{equation}
\Gamma_u=\Gamma_u^{(0)}+\left[(\Gamma_u^{{\rm kin}(1)}+\Gamma_u^{{\rm had}(1)})+
(\Gamma_u^{{\rm kin}(2)}+\Gamma_u^{{\rm had}(2)})\right]_{\rm BLNP}.
\end{equation}
The power corrections from the BLNP analysis 
are split into hadronic contributions, involving
subleading shape functions \cite{Lee:2004ja, Bosch:2004cb, Beneke:2004in}, 
and kinematic contributions,
which account for terms  suppressed by powers of
$P_+/M_B\sim P_+/P_-$ and thus scaling as $\Lambda_{\rm QCD}/M_B$ in
the shape-function region\footnote{ For a recent analysis
of these corrections in terms of subleading jet functions in SCET, see \cite{Paz:2009ut}.}.  The hadronic terms are treated at 
tree level, and the kinematic terms at one loop.

We summarize our results in the tables of numbers below, for the 
three partial rates studied in the previous section, as well as for
different values of the $E_l$ and $M_X$ cuts.   
The central values  and errors are obtained as follows. 
First, we evaluate the leading-power term 
$\Gamma_u^{(0)}$ as in Section~\ref{subsec:NNLOrates}.  We 
choose the default scales as $\mu_h^{\rm def}=4.25$~GeV and 
$\mu^{\rm def}_i=2.0$~GeV, and the HQET parameters as 
$m_b^*=4.707$~GeV and  $\mu_\pi^{*2}=0.216$~GeV$^2$.  Then, we evaluate
the power-suppressed terms {\it exactly} as in \cite{Lange:2005yw},
for the same choice of HQET parameters as for the leading term.  
Adding these two numbers together gives a default value for $\Gamma_u$.  
To this default value we assign uncertainties coming from a number of 
different sources.  To associate uncertainties with the HQET parameters, 
we vary them  simultaneously in the leading-order term and power corrections,
and assign errors for the ranges $m_b^*=\left(4.707^{+0.059}_{-0.053}\right)$~GeV
and $\mu_\pi^{*2}=\left(0.216^{+0.054}_{-0.076}\right)$~GeV$^2$.  
To estimate perturbative uncertainties associated with the choice of 
matching scales in the leading term,  
we vary the default choices for $\mu_h$ and 
$\mu_i$ up and  down by a factor of $\sqrt{2}$, and add these in quadrature
to obtain the uncertainty labeled $\mu_{h,i}$  in the tables.  Perturbative
scales also appear in the power corrections, through resummation
factors, and through the scale $\bar \mu$ for the kinematic power
corrections.  We vary these as in  \cite{Lange:2005yw}, and add
the different sources of perturbative uncertainty
in quadrature to obtain the total perturbative uncertainty for the 
power corrections, labeled $\mu_{\rm pow}$ in the tables.  Finally, we 
associate errors with the functional variation of the subleading
shape functions, called SSF in the tables, and add an uncertainty of 
$\delta\Gamma_u^{\rm WA}=\pm 1.26|V_{ub}|^2{\rm ps}^{-1}$
to take account weak annihilation effects (not shown in the table).
The analysis shows that the uncertainty from  $m_b^*$ is the 
dominant one. To quote the uncertainties in such a way that isolates this 
effect, we add all of the others in quadrature to obtain a
total uncertainty {\it excluding}  that associated with $m_b^*$, which
we instead quote in the last column of the tables.  The results read 
(in units of $|V_{ub}|^2 \, {\rm ps}^{-1}$):
\vspace{0.1cm}
\begin{table*}[t]
\begin{center}
\begin{tabular}{|lrcccc|}
\hline
Method & $\Delta{\cal B}^{\rm exp}~[10^{-4}]$~~ && $|V_{ub}|~[10^{-3}]$ 
 && $|V_{ub}|~[10^{-3}]$  \\
 &&& NLO && NNLO \\
\hline\\
[-0.4cm]
$E_l>2.1$\,GeV  & $3.3\pm 0.2\pm 0.7$
&& $3.56 \pm 0.40{}^{+0.48}_{-0.27}{}^{+0.31}_{-0.26}$ 
&&  $3.81\pm 0.43{}^{+0.33}_{-0.21}{}^{+0.31}_{-0.26}$ \\
CLEO \cite{Bornheim:2002du} &&&&& \\
$E_l>2.0$\,GeV    & $5.7\pm 0.4\pm 0.5$
&&  $3.97 \pm 0.22{}^{+0.37}_{-0.23}{}^{+0.26}_{-0.25}$ 
&&  $4.30 \pm 0.24{}^{+0.26}_{-0.20}{}^{+0.28}_{-0.27}$ \\
BABAR \cite{Aubert:2005mg}  &&&&& \\
$E_l>1.9$\,GeV  & $8.5\pm 0.4 \pm 1.5$
&&  $4.27 \pm 0.39{}^{+0.32}_{-0.19}{}^{+0.25}_{-0.22}$ 
&&  $4.65 \pm 0.43{}^{+0.27}_{-0.18}{}^{+0.27}_{-0.24}$ \\
BELLE \cite{Limosani:2005pi} &&&&& \\
$M_X<1.7$\,GeV   & $12.3\pm 1.1 \pm 1.2$
&&  $3.55 \pm 0.24{}^{+0.22}_{-0.13}{}^{+0.21}_{-0.19}$ 
&&  $3.87 \pm 0.26{}^{+0.21}_{-0.13}{}^{+0.21}_{-0.19}$\\
BELLE \cite{Bizjak:2005hn}  &&&&& \\
$M_X<1.55$\,GeV  & $11.7 \pm 0.9 \pm 0.7$
&&  $3.67 \pm 0.18{}^{+0.29}_{-0.17}{}^{+0.26}_{-0.24}$ 
&&  $3.96 \pm 0.19{}^{+0.20}_{-0.13}{}^{+0.26}_{-0.24}$ \\
BABAR \cite{Aubert:2007rb} &&&&& \\
$P_+<0.66$\,GeV  & $11.0\pm 1.0 \pm 1.6$ 
&&  $3.56 \pm 0.31{}^{+0.30}_{-0.17}{}^{+0.27}_{-0.23}$ 
&&  $3.84 \pm 0.33{}^{+0.21}_{-0.13}{}^{+0.26}_{-0.22}$ \\
BELLE \cite{Bizjak:2005hn} &&&&& \\
$P_+<0.66$\,GeV  & $9.4 \pm 1.0 \pm 0.8$
&&  $3.30 \pm 0.23{}^{+0.27}_{-0.16}{}^{+0.25}_{-0.22}$ 
&&  $3.55 \pm 0.24{}^{+0.19}_{-0.13}{}^{+0.24}_{-0.21}$ \\
BABAR \cite{Aubert:2007rb} &&&&& \\
\hline
\end{tabular}
\end{center}
\caption{\label{tab:Vub}
Values of $|V_{ub}|$ determined at NLO and NNLO, for the 
parameter values discussed in the text.  
The uncertainties  in the experimental measurements of 
$\Delta{\cal B}^{\rm exp}$ are statistical and systematic, respectively.
In the columns labeled $|V_{ub}|$
the first error is experimental, the second is the sum of all 
theoretical and parametric errors {\it except} for that from 
$m_b^*$, and the third  is that from $m_b^*$. We have combined errors 
of the same type  by adding in quadrature, and used 
$\tau_B = 1.584$~ps for the average $B$-meson lifetime. }
\end{table*}

\begin{itemize}

\item $P_+<0.66$~GeV:

\vspace{0.4cm}
\begin{tabular}{|c|c|c|c|c|c|c||c|}
\hline
 & $\Gamma_u$ & $\mu_{h,i}$ & $\mu_{\rm pow}$  &  $\mu_{\pi}^{*2}$ & SSF & tot   & $m_b^*$   \\ \hline \hline
NLO & 54.28 & {\footnotesize $^{+5.07}_{-7.36}$} & {\footnotesize $^{+1.91}_{-1.48}$} 
 &  {\footnotesize $^{+1.27}_{-1.07}$} 
 & {\footnotesize $\pm 1.41 $} & {\footnotesize $^{+5.87}_{-7.81}$}
& {\footnotesize $^{+8.20}_{-7.05}$} \\
NNLO & 46.77  & {\footnotesize $^{+1.44}_{-3.29}$} & {\footnotesize $^{+1.91}_{-1.48}$}
 &  {\footnotesize $^{+2.08}_{-1.56}$} 
&  {\footnotesize $\pm 1.41 $}  & {\footnotesize $^{+3.69}_{-4.36}$} 
&{\footnotesize $^{+6.34}_{-5.48}$} \\
\hline 
\end{tabular}

\item $M_X<1.55$~GeV:

\vspace{0.4cm}
\begin{tabular}{|c|c|c|c|c|c|c||c|}
\hline
 & $\Gamma_u$ & $\mu_{h,i}$ & $\mu_{\rm pow}$   & $\mu_{\pi}^{*2}$ & SSF & tot & $m_b^*$ \\ \hline \hline
NLO & 54.73 & {\footnotesize $^{+4.68}_{-7.08}$} & {\footnotesize $^{+1.99}_{-1.31}$} 
 & {\footnotesize $^{+1.02}_{-0.94}$} 
 & {\footnotesize $\pm 1.30 $} & {\footnotesize $^{+5.50}_{-7.48}$}  & {\footnotesize $^{+8.02}_{-6.87}$}
\\
NNLO & 47.09  & {\footnotesize $^{+1.22}_{-3.44}$} & {\footnotesize $^{+1.99}_{-1.31}$}
 & {\footnotesize $^{+1.79}_{-1.42}$} 
&  {\footnotesize $\pm 1.30 $} & {\footnotesize $^{+3.46}_{-4.34}$}  & {\footnotesize $^{+6.26}_{-5.41}$} \\
\hline 
\end{tabular}

\item $M_X<1.7$~GeV:

\vspace{0.4cm}
\begin{tabular}{|c|c|c|c|c|c|c||c|}
\hline
 & $\Gamma_u$ & $\mu_{h,i}$ & $\mu_{\rm pow}$   & $\mu_{\pi}^{*2}$ & SSF & tot & $m_b^*$ \\ \hline \hline
NLO & 61.21 & {\footnotesize $^{+3.69}_{-6.29}$} & {\footnotesize $^{+2.48}_{-1.60}$} 
 & {\footnotesize $^{+1.85}_{-1.46}$} 
 & {\footnotesize $\pm 0.63 $} & {\footnotesize $^{+5.02}_{-6.80}$}  & {\footnotesize $^{+7.36}_{-6.42}$}
\\
NNLO & 51.60  & {\footnotesize $^{+1.24}_{-4.13}$} & {\footnotesize $^{+2.48}_{-1.60}$}
 & {\footnotesize $^{+2.28}_{-1.72}$} 
&  {\footnotesize $\pm 0.63 $} & {\footnotesize $^{+3.85}_{-4.95}$}  & {\footnotesize $^{+5.80}_{-5.05}$} \\
\hline 
\end{tabular}

\newpage

\item $E_l>1.9$~GeV:

\vspace{0.4cm}
\begin{tabular}{|c|c|c|c|c|c|c||c|}
\hline
 & $\Gamma_u$ & $\mu_{h,i}$ & $\mu_{\rm pow}$   & $\mu_{\pi}^{*2}$ & SSF& tot & $m_b^*$ \\ \hline \hline
NLO & 29.21 & {\footnotesize $^{+2.14}_{-3.29}$} & {\footnotesize $^{+1.52}_{-1.01}$} 
  & {\footnotesize $^{+0.63}_{-0.51}$} 
 & {\footnotesize $\pm 0.53 $}  & {\footnotesize $^{+3.03}_{-3.73}$}   & {\footnotesize $^{+3.51}_{-2.91}$}
\\
NNLO & 24.65  & {\footnotesize $^{+0.63}_{-1.61}$} & {\footnotesize $^{+1.52}_{-1.01}$}
 & {\footnotesize $^{+0.75}_{-0.60}$} 
&  {\footnotesize $\pm 0.53 $}  & {\footnotesize $^{+2.27}_{-2.41}$}   & {\footnotesize $^{+2.95}_{-2.43}$}  \\
\hline 
\end{tabular}

\item $E_l>2.0$~GeV:

\vspace{0.4cm}
\begin{tabular}{|c|c|c|c|c|c|c||c|}
\hline
 & $\Gamma_u$ & $\mu_{h,i}$ & $\mu_{\rm pow}$   & $\mu_{\pi}^{*2}$ & SSF& tot & $m_b^*$ \\ \hline \hline
NLO & 22.76 & {\footnotesize $^{+2.34}_{-3.23}$} & {\footnotesize $^{+1.19}_{-0.88}$} 
  & {\footnotesize $^{+0.44}_{-0.35}$} 
 & {\footnotesize $\pm 0.59 $}  & {\footnotesize $^{+3.00}_{-3.64}$}   & {\footnotesize $^{+3.23}_{-2.66}$}
\\
NNLO & 19.38  & {\footnotesize $^{+0.63}_{-1.22}$} & {\footnotesize $^{+1.19}_{-0.88}$}
 & {\footnotesize $^{+0.60}_{-0.46}$} 
&  {\footnotesize $\pm 0.59 $}  & {\footnotesize $^{+2.03}_{-2.10}$}   & {\footnotesize $^{+2.70}_{-2.21}$}  \\
\hline 
\end{tabular}

\item $E_l>2.1$~GeV:

\vspace{0.4cm}
\begin{tabular}{|c|c|c|c|c|c|c||c|}
\hline
 & $\Gamma_u$ & $\mu_{h,i}$ & $\mu_{\rm pow}$   & $\mu_{\pi}^{*2}$ & SSF& tot & $m_b^*$ \\ \hline \hline
NLO & 16.30 & {\footnotesize $^{+2.50}_{-3.05}$} & {\footnotesize $^{+0.97}_{-0.89}$} 
  & {\footnotesize $^{+0.19}_{-0.16}$} 
 & {\footnotesize $\pm 0.70 $}  & {\footnotesize $^{+3.05}_{-3.49}$}   & 
{\footnotesize $^{+2.87}_{-2.31}$}
\\
NNLO & 14.17  & {\footnotesize $^{+0.64}_{-1.06}$} & {\footnotesize $^{+0.97}_{-0.89}$}
 & {\footnotesize $^{+0.38}_{-0.29}$} 
&  {\footnotesize $\pm 0.70 $}  & {\footnotesize $^{+1.89}_{-2.02}$}   
& {\footnotesize $^{+2.39}_{-1.93}$}  \\
\hline 
\end{tabular}
\end{itemize}

\vspace{0.4cm}
\noindent 
From these numbers and the corresponding experimental results, 
we arrive at the values of $|V_{ub}|$ listed in Table~\ref{tab:Vub}.
An examination of the table shows that the NNLO corrections shift
the values of $|V_{ub}|$ upwards by roughly 10\% compared to NLO.

\section{Conclusions}
\label{sec:Conclusions}
We studied the impact of NNLO perturbative
corrections on the leading term in the $1/m_b$ expansion for 
partial decay rates in $\bar B \to X_u l \bar \nu_l$ decays.
These corrections were implemented within a modified form of the BLNP
framework, which allows for  variations of both the perturbative jet scale
$\mu_i$ and the hard-matching scale $\mu_h$ in the resummed partial rates.
The particular choice $\mu_i=\mu_h$ corresponds to fixed-order 
perturbation theory, which  allowed us to perform a detailed comparison
between fixed-order results and the resummed results from BLNP.
Within resummed perturbation theory, we found that the dependence 
on the intermediate scale $\mu_i$ introduces sizeable 
perturbative uncertainties at NLO, which are reduced but still
significant at NNLO.  For the  conventional choice $\mu_i=1.5$~GeV used
in previous analyses in the BLNP formalism, the NNLO corrections also
induce fairly large downward shifts in the partial rates.  For higher values
of $\mu_i$ and in fixed-order perturbation theory, these shifts are 
more moderate.  We also compared between the full NNLO results in
fixed-order perturbation theory and those obtained in the 
large-$\beta_0$ approximation.  We found that the large-$\beta_0$
approximation for the NNLO corrections provides a poor approximation
to the full results, at least for the leading-order term in the 
$1/m_b$ expansion.  Whether this would also be true upon the inclusion
of terms suppressed by kinematic factors of $P_+/M_B$, 
which are small in the shape-function region, is an open question.
Finally, we combined our new results for the leading-order partial rates
with the known power corrections up to $1/m_b^2$, and showed how  
our analysis impacts the determination of $|V_{ub}|$ from several
experimental measurements. For parameter and scale choices typically
used in current analyses within the BLNP framework, the effect of 
the NNLO corrections is to raise the value of $|V_{ub}|$ by 
slightly less than 10\% compared to NLO; the exact results 
are shown in Table~{\ref{tab:Vub}.     

\vspace{0.4cm}

{\bf Acknowledgments}: 
C.G. is partially supported by the Swiss National Foundation as
well as EC-Contract MRTN-CT-2006-035482 (FLAVIAnet).
The Albert Einstein Center for Fundamental Physics (Bern) is
supported by the ``Innovations- und Kooperationsprojekt C-13 of
the Schweizerische Universit\"atskonferenz SUK/CRUS''.

\newpage

\begin{appendix}
\renewcommand{\theequation}{A\arabic{equation}}
\setcounter{equation}{0}

\section{Appendix}

\subsection{The hard and jet functions}
The hard functions $H_{ui}$ are derived from the SCET Wilson 
coefficients $C_i$,  which arise when 
matching the QCD $b\to u$ transition current 
onto SCET.  The two-loop QCD calculations needed for the NNLO
analysis were recently completed 
in \cite{Bonciani:2008wf, Asatrian:2008uk, Beneke:2008ei, Bell:2008ws}.
In terms of the quantities $H_{ij}=C_iC_j$, the  $H_{ui}$ 
read
\begin{eqnarray}
\label{eq:HuDefs}
H_{u1}=H_{11}, \qquad H_{u2}=0,  \qquad
H_{u3}= \frac{2 H_{13}+H_{33}}{y}+H_{12}+H_{23}+\frac{y}{4}H_{22} \,.
\end{eqnarray} 

The Laplace transformed jet function $\widetilde j$ was calculated
to NNLO in \cite{Becher:2006qw}.  To two-loop order, the result is 
\begin{equation}
   \widetilde j(L,\mu)
   = 1 + \frac{C_F\alpha_s}{4\pi} 
   \left( 2L^2 - 3L +7 - \frac{2\pi^2}{3} \right)
   + C_F \left( \frac{\alpha_s}{4\pi} \right)^2
   \left[ C_F J_F + C_A J_A + T_F n_f J_f \right] ,
\end{equation}
where
\begin{eqnarray}
   J_F &=& 2L^4 - 6L^3
    + \left( \frac{37}{2} - \frac{4\pi^2}{3} \right) L^2
    + \left( - \frac{45}{2} + 4\pi^2 - 24\zeta_3 \right) L
    + \frac{205}{8} - \frac{97\pi^2}{12} + \frac{61\pi^4}{90} - 6\zeta_3 \,,
    \nonumber\\
   J_A &=& - \frac{22}{9}\,L^3
    + \left( \frac{367}{18} - \frac{2\pi^2}{3} \right) L^2
    + \left( - \frac{3155}{54} + \frac{11\pi^2}{9} + 40\zeta_3 \right) L
    \nonumber\\
   &&\mbox{}+ \frac{53129}{648} - \frac{155\pi^2}{36} - \frac{37\pi^4}{180}
    - 18\zeta_3 \,, \nonumber\\
   J_f &=& \frac89\,L^3 - \frac{58}{9}\,L^2 
    + \left( \frac{494}{27} - \frac{4\pi^2}{9} \right) L
    - \frac{4057}{162} + \frac{13\pi^2}{9} \,.
\end{eqnarray}

\subsection{Renormalization-group factors and anomalous dimensions}
Here we list the perturbative expansion of the
renormalization-group functions in (\ref{eq:RGexps}) up to NNLO.  To do this,
we first define the expansion coefficients of the cusp anomalous dimension
and QCD $\beta$-function as 
\begin{eqnarray}
   \Gamma_{\rm cusp}(\alpha_s) &=& \Gamma_0\,\frac{\alpha_s}{4\pi}
    + \Gamma_1 \left( \frac{\alpha_s}{4\pi} \right)^2
    + \Gamma_2 \left( \frac{\alpha_s}{4\pi} \right)^3 
    + \Gamma_3 \left( \frac{\alpha_s}{4\pi} \right)^4 + \dots \,,
    \nonumber\\
   \beta(\alpha_s) &=& -2\alpha_s \left[ \beta_0\,\frac{\alpha_s}{4\pi}
    + \beta_1 \left( \frac{\alpha_s}{4\pi} \right)^2
    + \beta_2 \left( \frac{\alpha_s}{4\pi} \right)^3
    + \beta_3 \left( \frac{\alpha_s}{4\pi} \right)^4 + \dots \right] ,
\end{eqnarray}
and similarly for the other anomalous dimensions. In terms of these 
quantities, the function $a_\Gamma$ is given by \cite{Neubert:2004dd, Becher:2006mr} 
\begin{eqnarray}\label{asol}
   a_\Gamma(\nu,\mu)
   &=& \frac{\Gamma_0}{2\beta_0}\,\Bigg\{
    \ln\frac{\alpha_s(\mu)}{\alpha_s(\nu)}
    + \left( \frac{\Gamma_1}{\Gamma_0} - \frac{\beta_1}{\beta_0} \right)
    \frac{\alpha_s(\mu) - \alpha_s(\nu)}{4\pi} \nonumber\\ 
   &&\mbox{}+ \left[ \frac{\Gamma_2}{\Gamma_0} - \frac{\beta_2}{\beta_0}
    - \frac{\beta_1}{\beta_0}
    \left( \frac{\Gamma_1}{\Gamma_0} - \frac{\beta_1}{\beta_0} \right) \right]
    \frac{\alpha_s^2(\mu) - \alpha_s^2(\nu)}{32\pi^2} + \dots \Bigg\} \,.
\end{eqnarray}
Note that this result involves the three-loop anomalous dimension, which is 
known for the cusp anomalous dimension and for  $a_{\gamma^J}$, but not 
for the anomalous  dimension  $a_{\gamma'}$.
The result for the Sudakov factor $S$ is \cite{Becher:2006mr}
\begin{eqnarray}
   S(\nu,\mu) &=& \frac{\Gamma_0}{4\beta_0^2}\,\Bigg\{
    \frac{4\pi}{\alpha_s(\nu)} \left( 1 - \frac{1}{r} - \ln r \right)
    + \left( \frac{\Gamma_1}{\Gamma_0} - \frac{\beta_1}{\beta_0}
    \right) (1-r+\ln r) + \frac{\beta_1}{2\beta_0} \ln^2 r \nonumber\\
   &&\mbox{}+ \frac{\alpha_s(\nu)}{4\pi} \Bigg[ 
    \left( \frac{\beta_1\Gamma_1}{\beta_0\Gamma_0} - \frac{\beta_2}{\beta_0} 
    \right) (1-r+r\ln r)
    + \left( \frac{\beta_1^2}{\beta_0^2} - \frac{\beta_2}{\beta_0} \right)
    (1-r)\ln r \nonumber\\
   &&\hspace{1.0cm}
    \mbox{}- \left( \frac{\beta_1^2}{\beta_0^2} - \frac{\beta_2}{\beta_0}
    - \frac{\beta_1\Gamma_1}{\beta_0\Gamma_0} + \frac{\Gamma_2}{\Gamma_0}
    \right) \frac{(1-r)^2}{2} \Bigg] \nonumber\\
   &&\mbox{}+ \left( \frac{\alpha_s(\nu)}{4\pi} \right)^2 \Bigg[
    \left( \frac{\beta_1\beta_2}{\beta_0^2} - \frac{\beta_1^3}{2\beta_0^3}
    - \frac{\beta_3}{2\beta_0} + \frac{\beta_1}{\beta_0}
    \left( \frac{\Gamma_2}{\Gamma_0} - \frac{\beta_2}{\beta_0}
    + \frac{\beta_1^2}{\beta_0^2} - \frac{\beta_1\Gamma_1}{\beta_0\Gamma_0}
    \right) \frac{r^2}{2} \right) \ln r \nonumber\\
   &&\hspace{1.0cm}
    \mbox{}+ \left( \frac{\Gamma_3}{\Gamma_0} - \frac{\beta_3}{\beta_0}
    + \frac{2\beta_1\beta_2}{\beta_0^2} + \frac{\beta_1^2}{\beta_0^2}
    \left( \frac{\Gamma_1}{\Gamma_0} - \frac{\beta_1}{\beta_0} \right)
    - \frac{\beta_2\Gamma_1}{\beta_0\Gamma_0}
    - \frac{\beta_1\Gamma_2}{\beta_0\Gamma_0} \right) \frac{(1-r)^3}{3}
    \nonumber\\
   &&\hspace{1.0cm}
    \mbox{}+ \left( \frac{3\beta_3}{4\beta_0} - \frac{\Gamma_3}{2\Gamma_0}
    + \frac{\beta_1^3}{\beta_0^3}
    - \frac{3\beta_1^2\Gamma_1}{4\beta_0^2\Gamma_0}
    + \frac{\beta_2\Gamma_1}{\beta_0\Gamma_0}
    + \frac{\beta_1\Gamma_2}{4\beta_0\Gamma_0}
    - \frac{7\beta_1\beta_2}{4\beta_0^2} \right) (1-r)^2 \nonumber\\
   &&\hspace{1.0cm}
    \mbox{}+ \left( \frac{\beta_1\beta_2}{\beta_0^2} - \frac{\beta_3}{\beta_0}
    - \frac{\beta_1^2\Gamma_1}{\beta_0^2\Gamma_0}
    + \frac{\beta_1\Gamma_2}{\beta_0\Gamma_0} \right) \frac{1-r}{2}
    \Bigg] + \dots \Bigg\} \,,
\end{eqnarray}
where $r=\alpha_s(\mu)/\alpha_s(\nu)$. Whereas the three-loop anomalous 
dimensions and $\beta$-function are required in (\ref{asol}), the expression 
for $S$ also involves the four-loop coefficients $\Gamma_3$ and $\beta_3$.

We now list expressions for the anomalous dimensions and the QCD 
$\beta$-function, quoting all results in the $\overline{{\rm MS}}$ 
renormalization scheme.  The expansion of the cusp anomalous dimension
$\Gamma_{\rm cusp}$ to two-loop order was obtained some time ago 
\cite{Korchemskaya:1992je}, while recently the three-loop coefficient has been 
obtained in \cite{Moch:2004pa}. For the four-loop coefficient $\Gamma_3$, we 
follow \cite{Moch:2005ba} and use its [1,1] Pad\'e approximant,
$\Gamma_3\approx \Gamma_2^2/\Gamma_1$.  
The results are
\begin{eqnarray}
   \Gamma_0 &=& 4 C_F \,, \nonumber\\
   \Gamma_1 &=& 4 C_F \left[ \left( \frac{67}{9} - \frac{\pi^2}{3} \right)
    C_A - \frac{20}{9}\,T_F n_f \right] 
     \,, \nonumber\\
   \Gamma_2 &=& 4 C_F \Bigg[ C_A^2 \left( \frac{245}{6} - \frac{134\pi^2}{27}
    + \frac{11\pi^4}{45} + \frac{22}{3}\,\zeta_3 \right) 
    + C_A T_F n_f  \left( - \frac{418}{27} + \frac{40\pi^2}{27}
    - \frac{56}{3}\,\zeta_3 \right) \nonumber\\
   &&\mbox{}+ C_F T_F n_f \left( - \frac{55}{3} + 16\zeta_3 \right) 
    - \frac{16}{27}\,T_F^2 n_f^2 \Bigg] 
     \,, \nonumber\\
   \Gamma_3 &\approx& 7849,~ 4313, ~ 1553 \quad \mbox{for} \quad
    n_f = 3,\,4,\,5 \,.
\end{eqnarray}

The SCET anomalous dimension $\gamma'$ was deduced at two loops
using RG-invariance along with results for the  
jet and shape function anomalous dimensions in
\cite{Neubert:2004dd}, and was confirmed through 
the explicit calculations in  
\cite{Bonciani:2008wf, Asatrian:2008uk, Beneke:2008ei, Bell:2008ws} .  
The result is 
\begin{eqnarray}
   \gamma_0' &=& -5 C_F \,, \\
   \gamma_1' &=& -8 C_F \bigg[ \left(\frac{3}{16} - \frac{\pi^2}{4}
    + 3\zeta_3 \right) C_F
    + \left( \frac{1549}{432} + \frac{7\pi^2}{48} - \frac{11}{4}\,\zeta_3
    \right)\,C_A
    - \left( \frac{125}{216} + \frac{\pi^2}{24} \right) n_f \bigg] \,.
    \nonumber
\end{eqnarray}
To evaluate the RG-factor $a_{\gamma '}$ at NNLO requires the three-loop
anomalous dimension, which is not yet known. 
For that, we use the [1,1] Pad\'e approximation,
$\gamma_2'\approx \gamma_1'^2/\gamma_0'$.  Although this same approximation
works well for $\Gamma_{\rm cusp}$ it works poorly for $\gamma_J$,
as can be verified using the explicit results listed in \cite{Becher:2006mr}: 
\begin{eqnarray}\label{eq:gammaJ}
   \gamma_0^J &=& -3 C_F  \,, \nonumber\\
   \gamma_1^J &=& C_F^2 \left( - \frac{3}{2} + 2\pi^2 - 24\zeta_3 \right) 
    + C_F C_A \left( - \frac{1769}{54} - \frac{11\pi^2}{9} + 40\zeta_3 \right)
    + C_F T_F n_f \left( \frac{242}{27} + \frac{4\pi^2}{9} \right)
     \nonumber\\
   \gamma_2^J
   &=& C_F^3 \left( - \frac{29}{2} - 3\pi^2 - \frac{8\pi^4}{5} - 68\zeta_3 
    + \frac{16\pi^2}{3}\,\zeta_3 + 240\zeta_5 \right) \nonumber\\
   &&\mbox{}+ C_F^2 C_A \left( - \frac{151}{4} + \frac{205\pi^2}{9}
    + \frac{247\pi^4}{135} - \frac{844}{3}\,\zeta_3
    - \frac{8\pi^2}{3}\,\zeta_3 - 120\zeta_5 \right) \nonumber\\
   &&\mbox{}+ C_F C_A^2 \left( - \frac{412907}{2916} - \frac{419\pi^2}{243}
    - \frac{19\pi^4}{10} + \frac{5500}{9}\,\zeta_3
    - \frac{88\pi^2}{9}\,\zeta_3 - 232\zeta_5 \right) \nonumber\\
   &&\mbox{}+ C_F^2 T_F n_f \left( \frac{4664}{27} - \frac{32\pi^2}{9}
    - \frac{164\pi^4}{135} + \frac{208}{9}\,\zeta_3 \right) \nonumber\\
   &&\mbox{}+ C_F C_A T_F n_f \left( - \frac{5476}{729}
    + \frac{1180\pi^2}{243} + \frac{46\pi^4}{45}
    - \frac{2656}{27}\,\zeta_3 \right) \nonumber\\
   &&\mbox{}+ C_F T_F^2 n_f^2 \left( \frac{13828}{729} - \frac{80\pi^2}{81}
    - \frac{256}{27}\,\zeta_3 \right) \,
    \,.
\end{eqnarray}
Finally, the expansion coefficients for the QCD $\beta$-function to four-loop 
order are
\begin{eqnarray}
   \beta_0 &=& \frac{11}{3}\,C_A - \frac43\,T_F n_f  \,, \nonumber\\
   \beta_1 &=& \frac{34}{3}\,C_A^2 - \frac{20}{3}\,C_A T_F n_f
    - 4 C_F T_F n_f \,, \\
   \beta_2 &=& \frac{2857}{54}\,C_A^3 + \left( 2 C_F^2
    - \frac{205}{9}\,C_F C_A - \frac{1415}{27}\,C_A^2 \right) T_F n_f
    + \left( \frac{44}{9}\,C_F + \frac{158}{27}\,C_A \right) T_F^2 n_f^2
    \nonumber\\
   \beta_3 &=& \frac{149753}{6} + 3564\zeta_3
    - \left( \frac{1078361}{162} + \frac{6508}{27}\,\zeta_3 \right) n_f
    + \left( \frac{50065}{162} + \frac{6472}{81}\,\zeta_3 \right) n_f^2
    + \frac{1093}{729}\,n_f^3  \,, \nonumber
\end{eqnarray}
where  $\beta_3$ is taken from \cite{vanRitbergen:1997va} and is evaluated
for $N_c=3$.

\end{appendix}


\newpage






\begin{thebibliography}{99}

\bibitem{Barberio:2008fa}
  E.~Barberio {\it et al.}  [Heavy Flavor Averaging Group],
  arXiv:0808.1297 [hep-ex].


\bibitem{Antonelli:2009ws}
  M.~Antonelli {\it et al.},
  arXiv:0907.5386 [hep-ph].

\bibitem{Neubert:2008cp}
  M.~Neubert,
  arXiv:0801.0675 [hep-ph].

\bibitem{Neubert:1993ch}
  M.~Neubert,
  Phys.\ Rev.\  D {\bf 49}, 3392 (1994)
  [arXiv:hep-ph/9311325].

 
\bibitem{Neubert:1993um}
  M.~Neubert,
  Phys.\ Rev.\  D {\bf 49}, 4623 (1994)
  [arXiv:hep-ph/9312311].

\bibitem{Bigi:1993ex}
  I.~I.~Y.~Bigi, M.~A.~Shifman, N.~G.~Uraltsev and A.~I.~Vainshtein,
  Int.\ J.\ Mod.\ Phys.\  A {\bf 9}, 2467 (1994)
  [arXiv:hep-ph/9312359].



\bibitem{Bosch:2004th}
  S.~W.~Bosch, B.~O.~Lange, M.~Neubert and G.~Paz,
  Nucl.\ Phys.\  B {\bf 699} (2004) 335
  [arXiv:hep-ph/0402094].

\bibitem{Lange:2005yw}
  B.~O.~Lange, M.~Neubert and G.~Paz,
  Phys.\ Rev.\  D {\bf 72} (2005) 073006
  [arXiv:hep-ph/0504071].

\bibitem{Gambino:2007rp}
  P.~Gambino, P.~Giordano, G.~Ossola and N.~Uraltsev,
  JHEP {\bf 0710} (2007) 058
  [arXiv:0707.2493 [hep-ph]].

\bibitem{Andersen:2005mj}
  J.~R.~Andersen and E.~Gardi,
  JHEP {\bf 0601} (2006) 097
  [arXiv:hep-ph/0509360].


\bibitem{Bauer:2000yr}
  C.~W.~Bauer, S.~Fleming, D.~Pirjol and I.~W.~Stewart,
  Phys.\ Rev.\  D {\bf 63}, 114020 (2001)
  [arXiv:hep-ph/0011336].

\bibitem{Bauer:2001yt}
  C.~W.~Bauer, D.~Pirjol and I.~W.~Stewart,
  Phys.\ Rev.\  D {\bf 65} (2002) 054022
  [arXiv:hep-ph/0109045].


\bibitem{Beneke:2002ph}
  M.~Beneke, A.~P.~Chapovsky, M.~Diehl and T.~Feldmann,
  Nucl.\ Phys.\  B {\bf 643}, 431 (2002)
  [arXiv:hep-ph/0206152].

\bibitem{Korchemsky:1994jb}
  G.~P.~Korchemsky and G.~Sterman,
  Phys.\ Lett.\  B {\bf 340}, 96 (1994)
  [arXiv:hep-ph/9407344].


\bibitem{Akhoury:1995fp}
  R.~Akhoury and I.~Z.~Rothstein,
  Phys.\ Rev.\  D {\bf 54} (1996) 2349
  [arXiv:hep-ph/9512303].

\bibitem{Bauer:2003pi}
  C.~W.~Bauer and A.~V.~Manohar,
  Phys.\ Rev.\  D {\bf 70} (2004) 034024
  [arXiv:hep-ph/0312109].


\bibitem{Neubert:2004dd}
  M.~Neubert,
  Eur.\ Phys.\ J.\  C {\bf 40}, 165 (2005)
  [arXiv:hep-ph/0408179].


\bibitem{Becher:2006nr}
  T.~Becher and M.~Neubert,
  Phys.\ Rev.\ Lett.\  {\bf 97} (2006) 082001
  [arXiv:hep-ph/0605050].

\bibitem{Becher:2006mr}
  T.~Becher, M.~Neubert and B.~D.~Pecjak,
  JHEP {\bf 0701} (2007) 076
  [arXiv:hep-ph/0607228].

\bibitem{Bonciani:2008wf}
  R.~Bonciani and A.~Ferroglia,
  JHEP {\bf 0811} (2008) 065
  [arXiv:0809.4687 [hep-ph]].

\bibitem{Asatrian:2008uk}
  H.~M.~Asatrian, C.~Greub and B.~D.~Pecjak,
  Phys.\ Rev.\  D {\bf 78}, 114028 (2008)
  [arXiv:0810.0987 [hep-ph]].

\bibitem{Beneke:2008ei}
  M.~Beneke, T.~Huber and X.~Q.~Li,
  Nucl.\ Phys.\  B {\bf 811}, 77 (2009)
  [arXiv:0810.1230 [hep-ph]].


\bibitem{Bell:2008ws}
  G.~Bell,
  Nucl.\ Phys.\  B {\bf 812}, 264 (2009)
  [arXiv:0810.5695 [hep-ph]].



\bibitem{Becher:2006qw}
  T.~Becher and M.~Neubert,
  Phys.\ Lett.\  B {\bf 637}, 251 (2006)
  [arXiv:hep-ph/0603140].


\bibitem{Neubert:2004sp}
  M.~Neubert,
  Phys.\ Lett.\  B {\bf 612}, 13 (2005)
  [arXiv:hep-ph/0412241].

\bibitem{Becher:2005pd}
  T.~Becher and M.~Neubert,
  Phys.\ Lett.\  B {\bf 633}, 739 (2006)
  [arXiv:hep-ph/0512208].


\bibitem{Ligeti:2008ac}
  Z.~Ligeti, I.~W.~Stewart and F.~J.~Tackmann,
  Phys.\ Rev.\  D {\bf 78} (2008) 114014
  [arXiv:0807.1926 [hep-ph]].


\bibitem{Boos:2005by}
  H.~Boos, T.~Feldmann, T.~Mannel and B.~D.~Pecjak,
  Phys.\ Rev.\  D {\bf 73} (2006) 036003
  [arXiv:hep-ph/0504005].


\bibitem{Boos:2005qx}
  H.~Boos, T.~Feldmann, T.~Mannel and B.~D.~Pecjak,
  JHEP {\bf 0605} (2006) 056
  [arXiv:hep-ph/0512157].


\bibitem{Maitre:2005uu}
  D.~Maitre,
  Comput.\ Phys.\ Commun.\  {\bf 174}, 222 (2006)
  [arXiv:hep-ph/0507152].

\bibitem{Gehrmann:2001pz}
  T.~Gehrmann and E.~Remiddi,
  Comput.\ Phys.\ Commun.\  {\bf 141} (2001) 296
  [arXiv:hep-ph/0107173].


\bibitem{Campanario:2008xu}
  F.~Campanario, M.~Luke and S.~Zuberi,
  arXiv:0811.1787 [hep-ph].

\bibitem{Aquila:2005hq}
  V.~Aquila, P.~Gambino, G.~Ridolfi and N.~Uraltsev,
  Nucl.\ Phys.\  B {\bf 719} (2005) 77
  [arXiv:hep-ph/0503083].



\bibitem{Lee:2004ja}
  K.~S.~M.~Lee and I.~W.~Stewart,
  Nucl.\ Phys.\  B {\bf 721} (2005) 325
  [arXiv:hep-ph/0409045].

\bibitem{Bosch:2004cb}
  S.~W.~Bosch, M.~Neubert and G.~Paz,
  JHEP {\bf 0411}, 073 (2004)
  [arXiv:hep-ph/0409115].

\bibitem{Beneke:2004in}
  M.~Beneke, F.~Campanario, T.~Mannel and B.~D.~Pecjak,
  JHEP {\bf 0506} (2005) 071
  [arXiv:hep-ph/0411395].




\bibitem{Paz:2009ut}
  G.~Paz,
  JHEP {\bf 0906} (2009) 083
  [arXiv:0903.3377 [hep-ph]].

\bibitem{Bornheim:2002du}
  A.~Bornheim {\it et al.}  [CLEO Collaboration],
  Phys.\ Rev.\ Lett.\  {\bf 88} (2002) 231803
  [arXiv:hep-ex/0202019].


\bibitem{Aubert:2005mg}
  B.~Aubert {\it et al.}  [BABAR Collaboration],
  Phys.\ Rev.\  D {\bf 73}, 012006 (2006)
  [arXiv:hep-ex/0509040].

\bibitem{Limosani:2005pi}
  A.~Limosani {\it et al.}  [Belle Collaboration],
  Phys.\ Lett.\  B {\bf 621}, 28 (2005)
  [arXiv:hep-ex/0504046].

\bibitem{Bizjak:2005hn}
  I.~Bizjak {\it et al.}  [Belle Collaboration],
  Phys.\ Rev.\ Lett.\  {\bf 95}, 241801 (2005)
  [arXiv:hep-ex/0505088].

\bibitem{Aubert:2007rb}
  B.~Aubert {\it et al.}  [BABAR Collaboration],
  Phys.\ Rev.\ Lett.\  {\bf 100}, 171802 (2008)
  [arXiv:0708.3702 [hep-ex]].


























\bibitem{Korchemskaya:1992je}
  I.~A.~Korchemskaya and G.~P.~Korchemsky,
  Phys.\ Lett.\ B {\bf 287}, 169 (1992).

\bibitem{Moch:2004pa}
  S.~Moch, J.~A.~M.~Vermaseren and A.~Vogt,
  Nucl.\ Phys.\ B {\bf 688}, 101 (2004)
  [hep-ph/0403192].


\bibitem{Moch:2005ba}
  S.~Moch, J.~A.~M.~Vermaseren and A.~Vogt,
  Nucl.\ Phys.\ B {\bf 726}, 317 (2005)
  [hep-ph/0506288].







\bibitem{vanRitbergen:1997va}
  T.~van Ritbergen, J.~A.~M.~Vermaseren and S.~A.~Larin,
  Phys.\ Lett.\ B {\bf 400}, 379 (1997)
  [hep-ph/9701390].





\end{thebibliography}
\end{document}